\documentclass[twocolumn,preprintnumbers,amsmath,amssymb]{revtex4}
\let\tilde=\widetilde
\usepackage[unicode]{hyperref}
\usepackage[caption=false]{subfig}
\usepackage{graphicx}
\usepackage{dcolumn}
\usepackage{bm}
\usepackage{color}
\usepackage{amsfonts}
\usepackage{mathtools}
\usepackage{hyperref}
\usepackage{enumitem}

\newcommand{\dd}{\mbox{d}}
\newcommand{\ee}{\mbox{e}}

\newcommand{\mfrac}[2]{\mbox{$\frac{#1}{#2}$}}
\newcommand{\rec}[1]{\mfrac{1}{#1}}

\newcommand{\beq}{\begin{equation}}
\newcommand{\eeq}[1]{\label{#1}\end{equation}}
\newcommand{\bea}{\begin{eqnarray}}
\newcommand{\eea}[1]{\label{#1}\end{eqnarray}}

\def\a{\alpha}

\def\g{\gamma}
\def\G{\Gamma}
\def\d{\delta}
\def\D{\Delta}
\def\e{\epsilon}

\def\h{\eta}
\def\th{\theta}

\def\l{\lambda}

\def\n{\nu}
\def\x{\xi}

\def\p{\pi}

\def\s{\sigma}

\def\t{\tau}

\def\f{\phi}
\def\F{\Phi}

\def\ps{\psi}
\def\Ps{\Psi}

\def\de{\partial}

\begin{document}

\title{Stochastic Fractal and Noether's Theorem}%

\author{Rakibur Rahman$^{1,2}$, Fahima Nowrin$^2$, M.~Shahnoor Rahman$^2$, Jonathan A.~D.~Wattis$^3$ and Md.~Kamrul Hassan$^2$}
\date{\today}

\affiliation{
$^1$Max-Planck-Institut f\"ur Gravitationsphysik, Am M\"uhlenberg 1, D-14476 Potsdam-Golm, Germany\\
$^2$University of Dhaka, Department of Physics,
Theoretical Physics Group, Dhaka 1000, Bangladesh\\
$^3$School of Mathematical Sciences, University of Nottingham,
University Park, Nottingham NG7 2RD, UK}

\begin{abstract}

We consider the binary fragmentation problem in which, at any breakup event, one of the daughter segments either survives with probability $p$ or disappears with probability $1\!-\!p$.
It describes a stochastic dyadic Cantor set that evolves in time, and eventually becomes a fractal. We investigate this phenomenon, through analytical methods and Monte Carlo simulation, for a generic class of models, where segment breakup points follow a symmetric beta distribution with shape parameter $\a$, which also determines the fragmentation rate.
For a fractal dimension $d_f$, we find that the $d_f$-th moment $M_{d_f}$ is a conserved quantity, independent of $p$ and $\a$. We use the idea of data collapse$-$a consequence of dynamical scaling symmetry$-$to demonstrate that the system exhibits self-similarity. In an attempt to connect the symmetry with the conserved quantity,
we reinterpret the fragmentation equation as the continuity equation of a Euclidean quantum-mechanical system. Surprisingly, the Noether charge corresponding to dynamical
scaling is trivial, while $M_{d_f}$ relates to a purely mathematical
symmetry: quantum-mechanical phase rotation in Euclidean time.

\medskip
\end{abstract}


\maketitle


\section{Introduction}\label{sec-intro}

Natural objects rarely have regular shapes and smooth edges: they most often come with sparsely-distributed constituents,
badly-twisted tips, or wildly-folded surfaces. Similar features appear in a variety of seemingly disparate systems
across many branches of science.
Objects with such irregularities have traditionally been considered as geometrical
monsters, since Euclidean geometry confines us only to integer dimensions.
It was not until 1975, when Mandelbrot introduced the notion of
\emph{fractals}~\cite{ref.mandelbrot1,ref.mandelbrot2},
that proper appreciation was given to objects other than integer-dimensional Euclidean ones.
Unlike an Euclidean object, a fractal lacks uniform density; its degree of inhomogeneity is measured by the
\emph{fractal dimension}. Still, the two kinds of objects have an important feature in common, and that is \emph{self-similarity}.
In textbooks, what we usually see are fractals that arise after applying some
deterministic rules to simple Euclidean objects.
In nature, however, fractals come into existence through random processes and
time evolution. In order to incorporate randomness/stochasticity and the notion of time, one may instead apply
probabilistic rules in a sequential manner. This results in the so-called \emph{stochastic fractals}
(see~\cite{ref.Santo} for a detailed exposition). Fractals that appear in natural sciences are
generally stochastic in character. Self-similarity in such fractals manifests, only statistically, through
\emph{dynamical scaling} symmetry~\cite{ref.Santo}.

The role of randomness and time in the formation of fractals could be understood by considering a stochastic version
of the \emph{dyadic Cantor set} (DCS)~\cite{ref.Pandit}. Note that in the DCS problem, at each step,
every segment splits into equal halves, followed by the likely disappearance of a daughter segment with probability
$1-p$. In its stochastic counterpart, on the other hand, only one segment may split into two at any step; it splits
preferentially w.r.t.~its size, at a random point. We will see that, for $0<p<1$, the stochastic system evolves in time
to become a fractal. The system is self-similar in the sense that its snapshot at a given time is similar to that at
any other time. Moreover, the dynamics of the process brings along a conservation law$-$quite nontrivial for $p\neq1$.

In this article, we explore of the stochastic DCS problem for a generic class of models, where a shape parameter $\a$ encodes in a symmetric beta distribution the degree
of randomness in choosing the fragment breakup points, and tunes the fragmentation rate as well.
Finding analytical solutions and algorithms for numerical simulation, with a perfect matching between the two, has always been a formidable task for non-uniform distributions ($\a\neq1$). Our
present work achieves this feat. Apart from demonstrating the fractal nature of the system,
we prove that the $d_f$-th moment of the size distribution, $M_{d_f}$, is a conserved quantity and independent of $p$ and $\a$, where $d_f$ is the fractal dimension. Using the idea of data-collapse, we also show that the snapshot of the system at a given time resembles exactly that at any other time; this is the hallmark of continuous self-similarity symmetry. We investigate whether self-similarity and the conserved quantity $M_{d_f}$ are connected via Noether's theorem. Interestingly, $M_{d_f}$ relates as a Noether's charge not to self-similarity, but to a quantum-mechanical phase rotation in Euclidean time. On the other hand, the charge corresponding to self-similarity turns out to be zero.

The rest of this article is organized as follows. Section~\ref{sec-formulation} formulates the problem in its continuum version: a variant of the binary fragmentation equation~\cite{ref.Ziff_McGrady,ref.redner2}. In Section~\ref{sec-gen}, we discuss scaling theory and give
analytical solutions for $\a=1;2;3$. Section~\ref{sec-num} contains an exact algorithm to simulate the model for generic values of $\a$ and $p$. Extensive Monte Carlo simulations corroborate all our analytical results: a fractal dimension, a conserved moment and dynamical scaling symmetry.
Noether's theorem is invoked in Section~\ref{sec-Noether} in order to explore the symmetry origin of the conserved quantity.
In Section~\ref{sec-conc}, we make concluding remarks and leave some open questions.

\section{Formulation of the Problem}\label{sec-formulation}

Analytical studies of the stochastic DCS problem can be made by considering the following variant of the well-known binary
fragmentation equation~\cite{ref.Ziff_McGrady,ref.redner2}:
\bea \de_{t}c(x,t)&=&-c(x,t)\int_0^x\!\dd{y}\,F(x-y,y)\nonumber\\
&&+(1+p)\!\int_x^\infty\!\dd{y}\,c(y,t)F(x,y-x),\eea{eq:binery}
where $c(x,t)$ is the distribution function at time $t$ of segment size $x$. The first term on the right-hand side
of Eq.~(\ref{eq:binery}) describes the loss of size-$x$ segments due to their splitting into smaller ones, while the
second term describes the gain from the splitting of larger segments. The function $F(x,y)$, called the
\emph{fragmentation kernel}, captures the details of how a size-$(x+y)$ parent segment splits into two segments of
sizes $x$ and $y$. In particular, its integral gives the fragmentation rate~\cite{ref.Ziff_McGrady}:
\beq a(x)\equiv\int_0^x\!\dd{y}\,F(x-y,y),\eeq{frag_rate_defined}
which itself depends on the size $x$ of the parent segment. Moreover, the kernel furnishes with a probability
distribution of the breakup-point location along a given segment length. Last but not the least, the factor $1+p$ in
the gain term of Eq.~(\ref{eq:binery}) implies that, at each breakup event, the probability of survival is unity for
one of daughter segments, while that for the other one is $p$ (\emph{i.e.}, the other segment may disappear
with probability $1-p$). It is the latter factor in which Eq.~(\ref{eq:binery}) differs$-$rather crucially$-$from
the usual fragmentation equation~\cite{ref.Ziff_McGrady}. As we will see, the stochastic system~(\ref{eq:binery})
evolves in time, and eventually becomes a fractal, only when $p\neq1$.

On physical grounds, the fragmentation kernel ought to be symmetric with respect to exchange of its arguments as well
as a homogeneous function, \emph{i.e.},
\beq F(x,y)=F(y,x),\qquad F(\l x,\l y) = \l^\D F(x,y),\eeq{F-properties}
for some scaling dimension $\D \in \mathbb{R}$. The fragmentation rate~(\ref{frag_rate_defined}) scales as:
$a(\l x)=\l^{\D+1}a(x)$; the corresponding scaling dimension $\D+1$ is assumed to be positive:
\beq \D > -1.\eeq{condition-scaling}
If this condition does not hold, there occurs a cascading process in which smaller segments
break up at increasingly rapid rates, a.k.a.~the shattering transition, resulting in length being lost to a phase of
zero-size fragments~\cite{ref.Ziff_McGrady_shattering}. We instead  would like to consider the case where the segment
size approaches zero asymptotically in time, respecting the scaling limit:
\beq t\rightarrow\infty,\quad x\rightarrow0,\quad \text{such that}\quad x^{\D+1}t=\text{fixed}.\eeq{scaling-limit}
Because $x^{\D+1}t$ is a dimensionless quantity, such a limit always exists as long as
inequality~(\ref{condition-scaling}) holds.

The limit~(\ref{scaling-limit}) is defined, however, up to a dynamical scaling; it leaves room for
transformations of the kind:
\beq x\rightarrow x'=\l\,x,\qquad t\rightarrow t'=\l^zt,\eeq{dscaling1}
for some $\l\in\mathbb{R}_+$, with a negative \emph{dynamical exponent}:
\beq z=-(\D+1)<0.\eeq{dscaling2}
In the long-time limit, what happens if the transformations~(\ref{dscaling1})--(\ref{dscaling2}) comprise a symmetry
of the binary fragmentation problem~(\ref{eq:binery})? Because symmetry transformations of an equation
maps one solution to another, the size distribution $c(x,t)$ should exhibit a scaling behavior:
\beq c(\l x,\l^zt)=\l^{z\th}c(x,t),\eeq{dscaling3}
for some $\th\in\mathbb{R}$, in the long-time limit. The existence of scaling solutions can be revealed through
\emph{data collapse}: the hallmark of self-similarity in stochastic processes.

In this article, we will focus on the \emph{generalized product kernel},
of scaling dimension $\D=2(\a-1)$, given by:
\beq F(x-y,y)=\frac{(x-y)^{\a-1}y^{\a-1}}{B(\a,\a)}\,,\qquad \a>\tfrac{1}{2},\eeq{eq:generalized_kernel}
where the normalization factor $B(\a,\a)$ is an Euler beta function. One can rewrite the
kernel~(\ref{eq:generalized_kernel}) as:
\beq F(x-y,y)=x^{2(\a-1)}f(y/x;\a,\a),\eeq{eq:generalized_kernel2}
to find that $f(y/x;\a,\a)$ is nothing but the probability density function of the symmetric beta
distribution, for $y/x\in[0,1]$, with the shape parameter $\a$. With the kernel~(\ref{eq:generalized_kernel}),
the fragmentation rate~(\ref{frag_rate_defined}) takes the simple form:
\beq a(x)=x^{2\a-1},\eeq{frag_rate}
while the dynamical exponent $z$ takes the value:
\beq z=-(2\a-1)<0.\eeq{z-new}
Apart from determining the power law for the fragmentation rate w.r.t.~segment size,
the shape parameter$-$as the name suggests$-$spells out the shape of the probability
distribution with which segments split preferentially in the center. For example,
$\a=1$ gives a uniform probability distribution for the breakup point on a given segment length,
$\a=2$ corresponds to a parabolic distribution, while the $\a=3$ distribution curve takes the shape of a Mexican hat, and so on.
The distribution becomes sharper and sharper around the midpoint as $\a$ grows, approaching the
delta-function kernel in the limit $\a\rightarrow\infty$.

Previous studies of the generalized product kernel include the uniform~\cite{ref.Ziff_McGrady} and the parabolic
distributions~\cite{ref.Ziff_McGrady_product} in the context of classical binary fragmentation problem with $p=1$.
For $0<p<1$, the uniform distribution has already been investigated in~\cite{ref.Pandit}, where it was shown$-$both
analytically and through simulation$-$that the stochastic DCS exhibits dynamical scaling and possesses
a fractal dimension equal to $p$. This article aims, among other things, at a nontrivial generalization of the
latter work to an arbitrary shape parameter $\a>\tfrac{1}{2}$.

\section{Analytical Studies \label{sec-gen}}

In this section, we take an analytical approach to the stochastic DCS problem. We start with noting that the
kernel~(\ref{eq:generalized_kernel}) reduces the fragmentation equation~(\ref{eq:binery}) to:
\bea \de_{t}c(x,t)&=&-x^{2\a-1}c(x,t)~~~~~~~~~~~~~~~~~~~~~~~~~~~~~~~~~~~~~~~\nonumber\\
&&+\tfrac{1+p}{B(\a,\a)}\,x^{\a-1}\!\int_x^\infty\!\dd{y}\,(y-x)^{\a-1}c(y,t).\eea{eq:2}
With any given initial size distribution $c(x,0)$, solving this equation for $c(x,t)$ is a formidable task for
generic values of $\a>\tfrac{1}{2}$. The scaling behavior in the long-time limit comes in handy in this regard.
It is instructive to consider the $n$-th moment of the size-distribution function, defined as the
following Mellin transform:
\beq M_n(t)\equiv\int_0^\infty\!\dd x\,x^nc(x,t),\eeq{eq:moment}
and find the solution for $M_n(t)$~\cite{ref.krapivsky_1992,ref.krapivsky_naim}. Below we will obtain a
power-law behavior of $M_n(t)$ for large time.

\subsection{Power-Law Exponents \label{expo-sec}}

\begin{figure}[bt]
\centering
\subfloat[]{\includegraphics[width=4.3 cm,
clip=true]{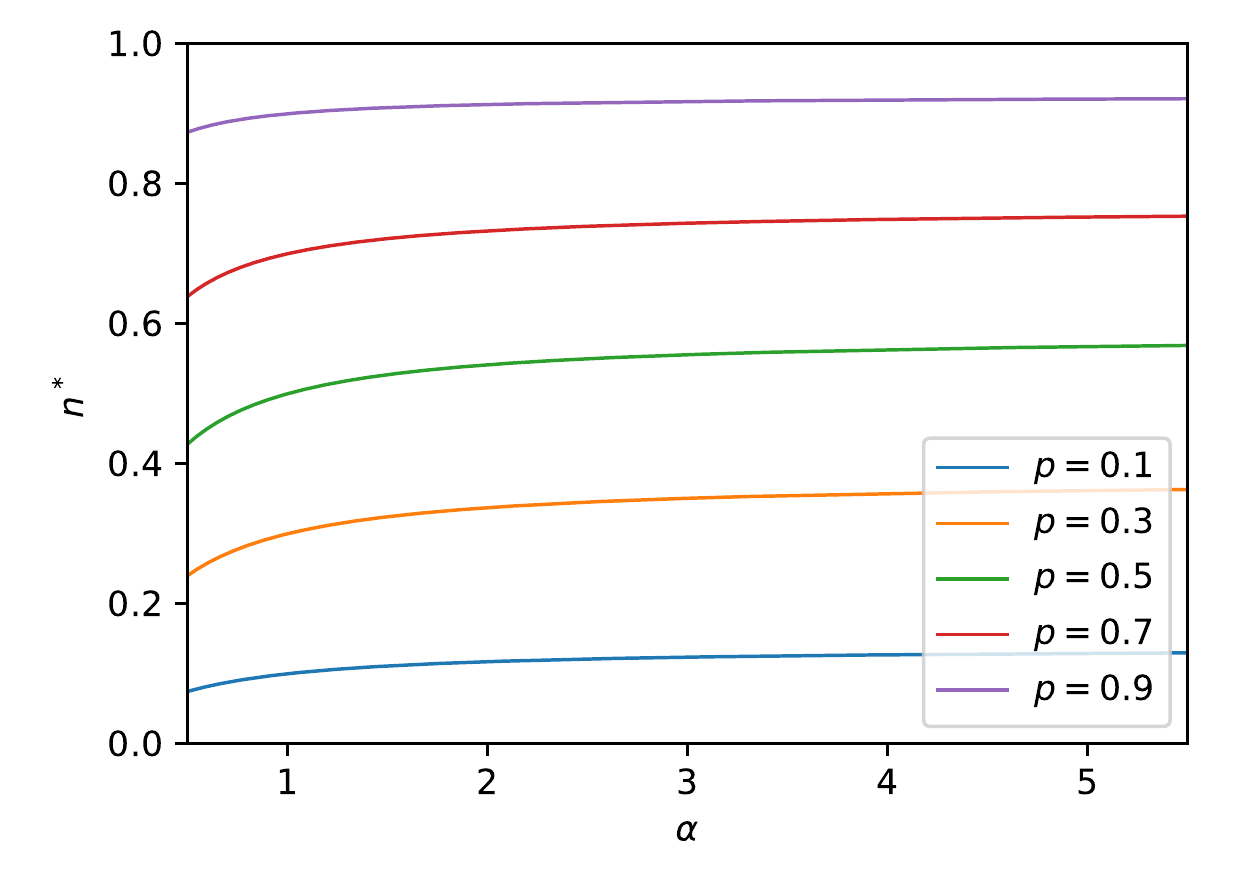} \label{fig:1a}}
\hspace{-10pt}
\subfloat[]{\includegraphics[width=4.3 cm,
clip=true]{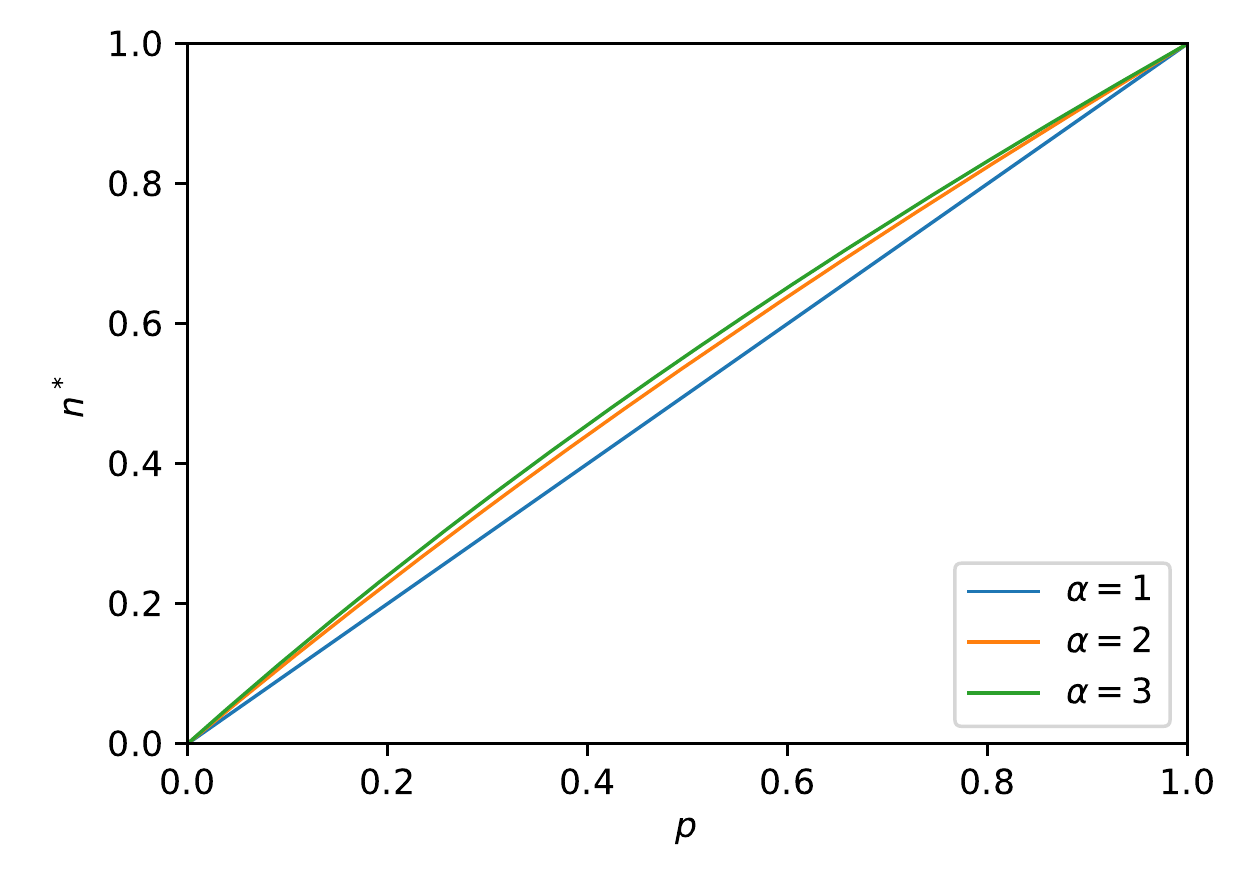} \label{fig:1b}}
\caption{Illustration of the dependency of $n^*$ on $\a$ and $p$.}
\label{fig:1ab}
\end{figure}

In order to incorporate the definition of $M_n(t)$, we take a Mellin transform of Eq.~(\ref{eq:2}),
and obtain:
\beq \frac{d}{dt}M_n(t)=\g_n\,M_{n+2\a-1}(t),\eeq{eq:rate_equation}
where $\g_n=\g_n(\a, p)$ has been defined as:
\beq \g_n(\a, p)\equiv\frac{(1+p)B(\a,\a+n)}{B(\a,\a)}-1.\eeq{eq:polynomial}
The transcendental equation $\g_{n}(\a, p)=0$ will always have a positive real root $n=n^*$, such that
$0\!<\!n^*\!<\!1$ for $p\in(0,1)$. To see this, it suffices to note that $\g_{n=0}=p>0$, while
$\g_{n=1}=-\tfrac{1}{2}(1-p)<0$. Therefore, $\g_n$ must vanish somewhere in the open interval $(0,1)$:
\beq \exists\,n^*\in(0,1)\quad \text{such that}\quad \g_{n^*}(\a, p)=0.\eeq{eq:polynomial2}
Then, Eq.~(\ref{eq:rate_equation}) tells us that there exists a moment of the size distribution that
is conserved in
time~\cite{ref.krapivsky_1992,ref.krapivsky_naim,ref.Hassan_Rodgers,ref.Hassan_Kurths}:
\beq \frac{d}{dt}M_{n^*}(t)=0,\eeq{eq:conserved_moment}
for some $n^*\in(0,1)$, whose value can be determined by solving the transcendental
equation~(\ref{eq:polynomial2}), \emph{i.e.},
\beq \frac{\Gamma(n^*+\a)\,\Gamma(2\a)}{\Gamma(n^*+2\a)\,\Gamma(\a)}=\frac{1}{1+p}\,.\eeq{eq:p-nstar}
While Eq.~(\ref{eq:p-nstar}) reduces, for integer values of $\a$, to a polynomial equation that could
be solved algebraically, the value of $n^*$ in the generic case could be obtained numerically.
Figure~\ref{fig:1ab} illustrates the behavior of $n^*$ as a function of $\a$ and $p$.
For a given $p$, as $\a$ grows, $n^*$ increases to reach the asymptotic value of $\log_2(1+p)$.
With $\a$ given, $n^*$ increases monotonically with $p$ towards unity.

The rate equation~(\ref{eq:rate_equation}) can help finding an expression for $M_n(t)$~\cite{ref.Charlesby}.
Consider the Taylor expansion of $M_n(t)$ around $t=0$; each term in the expansion can be computed by
iterating Eq.~(\ref{eq:rate_equation}). With a monodisperse initial condition: $c(x,0)=\d(x-l)$, for example,
one arrives at:
\beq M_n(t)=l^n\left[1+\g_n t+\tfrac{1}{2!}\g_n\g_{n+(2\a-1)}t^2+\cdots\right].\eeq{Mn_Taylor}
As expected, if $n=n^*$, the corresponding moment will not depend on time, since $\g_{n^*}$ vanishes.
Moreover, if $n$ is of the form: $n=n^*-k(2\a-1)$, for $k\in\mathbb{N}$, the corresponding series will
stop at $t^k$. In other words, in the long-time limit, one can write:
\beq M_n(t)\sim t^{-\left(\tfrac{n-n^*}{2\a-1}\right)}.\eeq{eq:moment_solution}
Although this result has been obtained only for $n$ being less than $n^*$ by integer multiples of $(2\a-1)$,
it actually holds good for any real value of $n$. This can be seen by recourse to the scaling behavior
that we consider next.

Given the large-time scaling behaviors~(\ref{dscaling1})--(\ref{dscaling3}), it follows from the Buckingham
$\Pi$ theorem that the scaling form of the size distribution can be written as:
\beq c(x,t)\sim t^\th\phi(xt^{-1/z}),\eeq{eq:dynamic_scaling}
for some $\th\in\mathbb{R}$, such that $c(x,t)/t^\th$ is dimensionless, just as
$xt^{-1/z}$ is~\cite{ref.Dongen,ref.barenblatt,ref.redner1}. The physical significance of the scaling
form is as follows. If snapshots of the system are taken at different time-points
to collect data for $c(x,t)$ as a function of $x$, then the resulting plots will be distinct for
each time-point. However, plots of the corresponding dimensionless
quantities, $c(x,t)/t^\th$ versus $xt^{-1/z}$, must collapse into one universal curve $\phi(xt^{-1/z})$,
known as the \emph{scaling function}. This phenomenon of data collapse is the hallmark of self-similarity
in stochastic processes. Self-similarity in this case means that the snapshots of the same system taken
at different times$-$although different$-$will look indistinguishable when appropriate temporal rescalings
(with exponents $z$ and $\th$) are made.

To determine the exponent $\th$, let us plug the scaling form~(\ref{eq:dynamic_scaling}) into the
definition~(\ref{eq:moment}), and obtain:
\beq M_{n}(t)\sim t^{\th+(n+1)/z}\int_0^\infty\!\dd{\x}\,\x^n \phi(\x),\eeq{power_law_theta}
where we have introduced the following denotation:
\beq\x\equiv xt^{-1/z}.\eeq{xi-defined}
Because $\x$ is dimensionless, the integral in Eq.~(\ref{power_law_theta}) is just an irrelevant numerical factor.
Now, we recall that $n^*$-th moment does not depend on time. Then, it follows that:
\beq \th\,=\,-\frac{n^*+1}{z}\,=\,\frac{n^*+1}{2\a-1}\,.\eeq{eq:th}
Finally, we substitute this expression for $\th$ back into Eq.~(\ref{power_law_theta}). The result is nothing but
our old equation~(\ref{eq:moment_solution}), now proved for an arbitrary real value of $n$.

The moments of the size distribution correspond to various physical quantities:
the zeroth moment $M_0(t)$ gives the total number of segments at time $t$, the first moment $M_1(t)$
measures the total length of the segments, the second moment $M_2(t)$ is related to variance of segment size,
\emph{etc.} The mean or typical segment size will be given by the ratio: $\d(t)\equiv M_1(t)/M_0(t)$.
According to Eq.~(\ref{eq:moment_solution}), the latter quantity decays in time as:
\beq \d(t)\sim t^{-1/(2\a-1)}.\eeq{eq:mean_particle}
In the expression for the zeroth moment resulting from Eq.~(\ref{eq:moment_solution}), if we eliminate $t$ in favor of $\d$, we would obtain:
\beq M_0(\d)\sim\d^{-n^*}.\eeq{eq:Ndf}
As is well known, the exponent of such a power-law relation is to be interpreted as the Hausdorff dimension $d_f$
of the system (see, for example,~\cite{ref.Santo}). Therefore, we have:
\beq d_f=n^*\in(0,1),\quad\text{for}\quad0<p<1.\eeq{eq:df_obtained}
Because it takes a non-integer value less than the dimension
of the embedding Euclidean space, we would identify the Hausdorff dimension
$d_f=n^*$ as the fractal dimension of the binary fragmentation problem.
On the other hand, as already discussed, self-similarity is ensured
by the very existence of the scaling limit~(\ref{scaling-limit}), with
$\D=2(\a-1)>-1$. These two$-$self-similarity and fractal dimensionality$-$confirm
the fractal nature of the system.

\subsection{Scaling Functions \label{dist-sec}}

The scaling form~(\ref{eq:dynamic_scaling}) of the size distribution contains a universal scaling function
$\f(\x)$ of the dimensionless variable $\xi$, defined in Eq.~(\ref{xi-defined}).
Now we would like to elaborate on the nature of the scaling function.
Finding an analytical solution for $\f(\x)$ is very difficult in general.
In our modest attempt, we would restrict the analyses to some convenient (small integer)
values of the shape parameter $\a$. With the end in view, let us substitute the scaling
form~(\ref{eq:dynamic_scaling}) into Eq.~(\ref{eq:2}). This leads to the following
integro-differential equation:
\bea 0&=&\x\f'(\x)+(2\a-1)(\x^{2\a-1}+\th)\f(\x)~~~~~~~~~~\nonumber\\
&&-(1+p)\,b(\a)\,\x^{\a-1}\!\int_{\h=\x}^\infty\!\dd{\h}\,
(\h-\x)^{\a-1}\f(\h),~~~~~\eea{ide}
where a ``prime'' denotes a derivative w.r.t.~$\xi$, and
\beq b(\a)\equiv\tfrac{2\a-1}{B(\a,\a)}\,.\eeq{b-defined}
Note that all the time-dependencies have been dropped in Eq.~(\ref{ide}) since they contribute to an overall factor,
thanks to the explicit values of the parameters $z$ and $\th$.

For integer values of $\a$, we already noted that Eq.~(\ref{eq:p-nstar}) reduces to a polynomial equation in $n^*$.
In this case, $p=p(n^*)$ will be a polynomial function, so that $\f(\x)$ could have closed-form solutions in terms of
generalized hypergeometric functions, as we will see. For given $\a$, it is more convenient to leave $n^*$ as the free parameter
in Eq.~(\ref{ide}). It is also useful to introduce a new function:
\beq \F(\x)\equiv\int_{\eta=\x}^\infty\!\dd{\h}\,\f(\h),\qquad \f(\x)=-\F'(\x),\eeq{newscaling}
which satisfies the following boundary conditions:
\beq \F(\x=0)=1,\qquad \F(\x\rightarrow\infty)=0.\eeq{newscaling_bc}
While the first boundary condition is just a choice of normalization, the second one follows
from the definition of $\F(\x)$.
We now present analytical solutions for the scaling function in the cases of $\a=1;2;3$.
Necessary expressions for the relevant parameters are summarized in Table~\ref{param-tab}.
\begin{table}
\begin{center}
\begin{tabular}{ c | c | c | c }
$\a$\;&\; $b(\a)$ [Eq.~(\ref{b-defined})]\;&\;$p$ [Eq.~(\ref{eq:p-nstar})]\;\;&\;\; $\th$ [Eq.~(\ref{eq:th})]\\ \hline
1 & $1$ & $n^*$ & $n^*\!+\!1$ \\
2 & $18$   & $\rec{6}n^*(n^*+5)$ & \; $\rec{3} (n^*\!+\!1)$ \\
3 & $150$ & \; $\rec{60}n^* (n^{*2} \!+\! 12 n^*\!+\!47)$
\; & \; $\rec{5}(n^*\!+\!1)$ \\
\end{tabular}
\end{center}
\caption{Parameter values for the special cases $\a=1;2;3$.
\label{param-tab}}
\end{table}
%

\subsubsection{Uniform Distribution: $\a=1$ \label{beta0-sol-sec}}

Already studied in~\cite{ref.Pandit}, this case is included here for the sake of completeness.
Using Table~\ref{param-tab}, Eq.~(\ref{ide}) can be simplified to the following form:
\beq \x\f'(\x)+(\x\!+\!n^*\!+\!1)\f(\x)
-(n^*\!+\!1)\!\int_{\eta=\x}^\infty\!\dd{\h}\,\f(\h)=0.\eeq{ide0}
In terms of $\F(\xi)$, this reduces to a differential equation:
\beq \x\F''(\x)+(\x\!+\!n^*\!+\!1)\F'(\x)+(n^*\!+\!1)\F(\x)=0,\eeq{ide001}
which has the general solution:
\beq \F(\x)=C_1\ee^{-\x}+C_2\ee^{-\x+i\p n^*}\G(-n^*,-\x).\eeq{ide002}
In view of the boundary conditions~(\ref{newscaling_bc}), we must have:
$C_1=1, C_2=0$. Finally, a differentiation w.r.t.~$\x$ gives:
\beq \f(\x)=\ee^{-\x},\eeq{bo-sol-final}
which, rather remarkably, is independent of $p$.

\subsubsection{Parabolic Distribution: $\a=2$ \label{beta1-sol-sec}}

In this case, given Table~\ref{param-tab}, Eq.~(\ref{ide}) reduces to:
\bea 0&=& \x\f'(\x)+(3\xi^3\!+\!n^*\!+\!1)\f(\x)\nonumber\\
&&-3(n^*\!+\!2)(n^*\!+\!3)\,\x\int_{\h=\x}^\infty\!\dd{\h}\,(\h-\x)\f(\h).
\eea{ide1}
Now, let us take a logarithmic derivative of the above equation, \emph{i.e.},
operate it by $\x\tfrac{d}{d\x}$. From the resulting equation, when Eq.~(\ref{ide1}) is subtracted,
there will be some cancellation among the integral terms. Thus, we end up with a third-order
differential equation for $\F(\x)$; it reads:
\bea 0&=&\x^2\F'''(\x)\!+\!(3\x^3\!+\!n^*\!+\!1)\x\F''(\x)\!+\!(6\x^3\!-\!n^*\!-\!1)\F'(\x)\nonumber\\
&&-3(n^*\!+\!2)(n^*\!+\!3)\,\x^2\,\F(\x),\eea{Q_diff1}
whose general solution takes the following form:
\beq \F(\x)=C_1\F_1(\x)+C_2\F_2(\x)+C_3\F_3(\x),\eeq{Qgen}
where the three different components are generalized hypergeometric functions (see, for example,~\cite{ref:olver}):
\bea &\F_1={}_1F_1\left( -\tfrac{n^*+2}{3}, \tfrac{1}{3} ; -\x^3\right),&\nonumber\\
&\F_2=\x^2{}_1F_1\left(-\tfrac{n^*}{3}, \tfrac{5}{3} ; -\x^3\right),&\label{Qgen-components}\\
&\F_3=\x^{-n^*}{}_2F_2\left( 1,-\tfrac{2(n^*+1)}{3} ; \tfrac{3-n^*}{3},\tfrac{1-n^*}{3} ;
-\x^3\right).&\nonumber\eea{xxyyzz}
Given the asymptotic forms of the generalized hypergeometric functions
for large $\x$, the vanishing of $\F(\x\rightarrow\infty)$ leaves us with a unique linear
combination of the component solutions~(\ref{Qgen-components}). In addition to that, the normalization
condition fixes the overall constant, giving:
\beq C_1=1,\qquad C_2=-\tfrac{\G\left(\tfrac{1}{3}\right)\G\left(\tfrac{n^*+5}{3}\right)}
{\G\left(\tfrac{5}{3}\right)\G\left(\tfrac{n^*+3}{3}\right)},\qquad C_3=0.\eeq{Qgen2}

\begin{figure}[bt]
\centering
\subfloat[]{\includegraphics[width=4.3 cm,
clip=true]{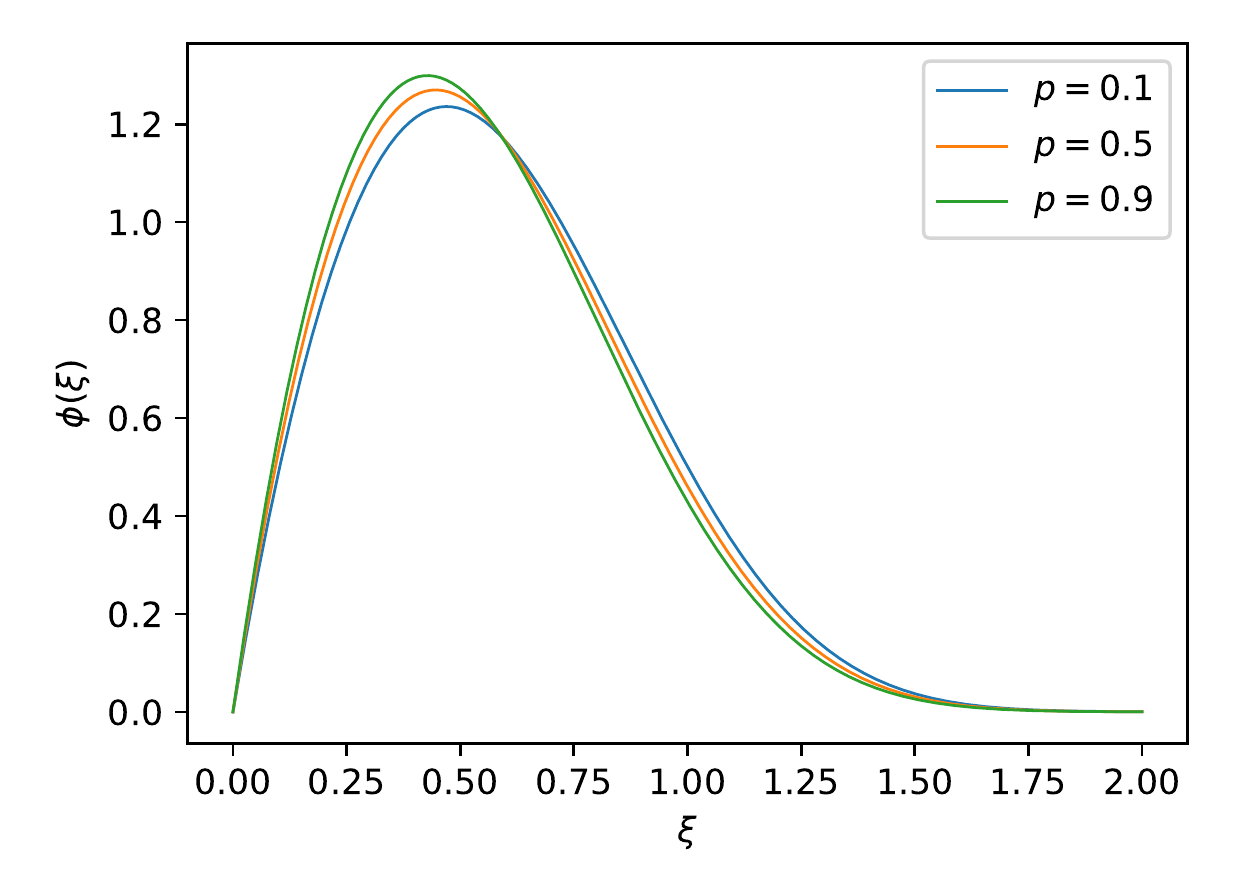} \label{fig:2a}}
\hspace{-10pt}
\subfloat[]{\includegraphics[width=4.3 cm,
clip=true]{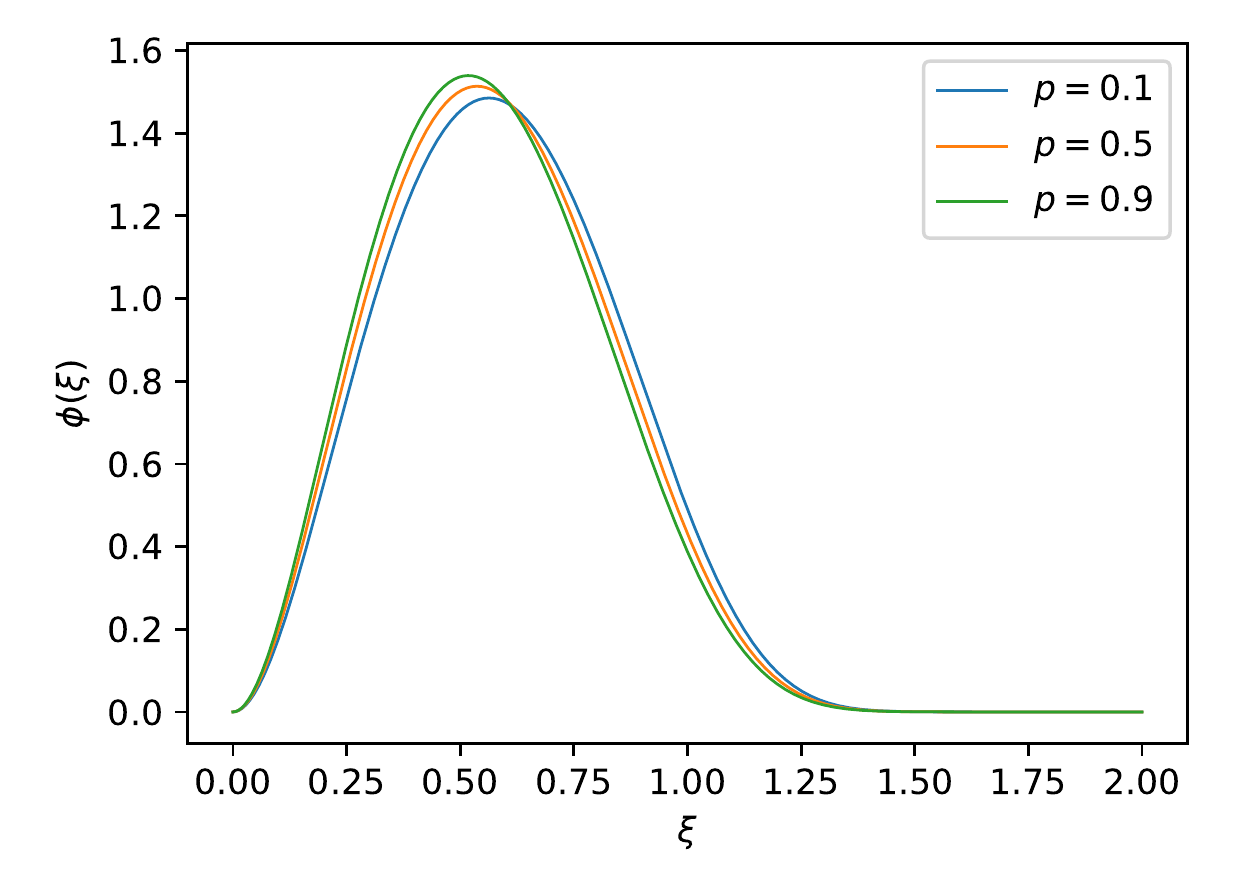} \label{fig:2b}}
\caption{Plots of the scaling function $\phi(\xi)$ for (a) parabolic distribution: $\a=2$, and (b) Mexican-hat distribution: $\a=3$.}
\label{fig:2ab}
\end{figure}

The explicit solution for $\f(\x)$ is then obtained by differentiating Eq.~(\ref{Qgen}) w.r.t.~$\x$;
the result is:
\bea \f(\x)&=&-3(n^*+2)\x^2\,{}_1F_1\left(-\tfrac{n^*-1}{3}, \tfrac{4}{3} ; -\x^3\right)\nonumber\\
&&-\tfrac{3}{5}C_2n^*\x^4\,{}_1F_1\left(-\tfrac{n^*-3}{3}, \tfrac{8}{3}; -\x^3\right)\label{solphia2}\\
&&-2C_2\x\,{}_1F_1\left(-\tfrac{n^*}{3}, \tfrac{5}{3} ; -\x^3\right),\nonumber\eea{xx1yy1zz1}
with $C_2$ given in Eq.~(\ref{Qgen2}). The solution~(\ref{solphia2}) depends on $p$$-$through the
relation: $p=\tfrac{1}{6}n^*(n^*+5)$$-$in a nontrivial manner. However, this dependency is rather weak,
as we see from Figure~\ref{fig:2a}, where we plot the scaling function~(\ref{solphia2}) for the cases
of $p=0.1, 0.5$, and $0.9$.  Therefore, we would not expect the $p$-dependency of $\f(\x)$ to be clearly noticeable
in numerical simulations.
%

\subsubsection{Mexican-Hat Distribution: $\a=3$ \label{beta2-sol-sec}}

In the case $\a=3$, the data from Table~\ref{param-tab}
enables the simplification of Eq.~(\ref{ide}) to:
\bea 0&=& \x\f'(\x)+(5\xi^5\!+\!n^*\!+\!1)\f(\x)\nonumber\\
&&-\tfrac{5(n^*+5)!}{2(n^*+2)!}\,\x^2\int_{\h=\x}^\infty\!\dd{\h}\,(\h-\x)^2\f(\h).\eea{ide31}
One can operate this equation by $\left(\xi^2\tfrac{d}{d\x^2}-4\x\tfrac{d}{d\x}+6\right)$
to obtain a fourth-order differential equation for $\F(\x)$:
\bea
0&=&\x^3\F''''(\x)+(5\x^5\!+\!n^*\!-\!1)\x^2\F'''(\x)\nonumber \\
 &&+(30\x^5\!-\!4n^*\!-\!2)\x\F''(\x)\!+\!6(5\x^5 \!+\!n^*\!+\!1)\F'(\x)~~~~~~~\label{Qode2}\\
 &&+5(n^*\!+\!3)(n^*\!+\!4)(n^*\!+\!5)\x^4\F(\x).\nonumber\eea{x2y2z2}
This equation has a general solution of the form:
\beq \F(\x)=C_1\F_1(\x)+C_2\F_2(\x)+C_3\F_3(\x)+C_4\F_4(\x),\eeq{Q2-sol}
with the four components given by:
\begin{equation}\label{Q2-sol-com} \begin{split}
\F_1&={}_2F_2\left(-\tfrac{2+\n}{10},-\tfrac{2+\bar{\n}}{10};\,\tfrac{1}{5},\tfrac{2}{5};\,-\x^5\right),\\
\F_2&=\x^3\,{}_2F_2\left(\tfrac{4-\n}{10},\tfrac{4-\bar{\n}}{10};\,\tfrac{4}{5},\tfrac{8}{5};\,-\x^5\right),\\
\F_3&=\x^4\,{}_2F_2\left(\tfrac{6-\n}{10},\tfrac{6-\bar{\n}}{10};\,\tfrac{6}{5},\tfrac{9}{5};\,-\x^5\right),\\
\F_4&=\x^{-n^*}\!{}_3F_3\!\left(\!1,\tfrac{\tilde{\n}}{10},\tfrac{\bar{\tilde{\n}}}{10};\tfrac{1-n^*}{5},\tfrac{2-n^*}{5},
\tfrac{5-n^*}{5};-\x^5\!\right),\end{split}\end{equation}
where $\n\equiv n^*\!+\!i\sqrt{3n^{*2}\!+\!24n^*\!+\!44}$ is a complex number, so is $\tilde{\n}\equiv-\n\!-\!2(n^*\!+\!1)$,
and a ``bar'' denotes a complex conjugation. Again, there is a unique linear combination of the components compatible with
the vanishing of $\F(\x\rightarrow\infty)$, while the overall factor is fixed by $\F(\x=0)$ being unity. The coefficients
are thus determined to be:
\beq C_1=1,\quad C_2=\tfrac{a_3b_1-a_1b_3}{a_2b_3-a_3b_2},\quad C_3=\tfrac{a_1b_2-a_2b_1}{a_2b_3-a_3b_2},\quad C_4=0,\eeq{c1c2c3_defined}
with the constants $a_i, b_i (i=1,2,3)$ given as:
\begin{equation}\label{Q2-sol-cons} \begin{split}
a_1&=\tfrac{\G\left(\tfrac{1}{5}\right)\,\G\left(\tfrac{2}{5}\right)\,\G\left(\tfrac{n^*-\n}{5}\right)}
{\G\left(\tfrac{-2-\n}{10}\right)\,\G\left(\tfrac{2n^*+4-\n}{10}\right)\,\G\left(\tfrac{2n^*+6-\n}{10}\right)}\,,\\
b_1&=\tfrac{\G\left(\tfrac{1}{5}\right)\,\G\left(\tfrac{2}{5}\right)\,\G\left(\tfrac{\n-n^*}{5}\right)}
{\G\left(\tfrac{4+\n}{10}\right)\,\G\left(\tfrac{6+\n}{10}\right)\,\G\left(\tfrac{\n-2n^*-2}{10}\right)}\,,\\
a_2&=\tfrac{\G\left(\tfrac{4}{5}\right)\,\G\left(\tfrac{8}{5}\right)\,\G\left(\tfrac{n^*-\n}{5}\right)}
{\G\left(\tfrac{4-\n}{10}\right)\,\G\left(\tfrac{2n^*+4-\n}{10}\right)\G\left(\tfrac{2n^*+12-\n}{10}\right)}\,,\\
b_2&=\tfrac{\G\left(\tfrac{4}{5}\right)\,\G\left(\tfrac{8}{5}\right)\,\G\left(\tfrac{\n-n^*}{5}\right)}
{\G\left(\tfrac{4+\n}{10}\right)\,\G\left(\tfrac{12+\n}{10}\right)\,\G\left(\tfrac{\n-2n^*+4}{10}\right)}\,,\\
a_3&=\tfrac{\G\left(\tfrac{6}{5}\right)\,\G\left(\tfrac{9}{5}\right)\,\G\left(\tfrac{n^*-\n}{5}\right)}
{\G\left(\tfrac{6-\n}{10}\right)\,\G\left(\tfrac{2n^*+6-\n}{10}\right)\,\G\left(\tfrac{2n^*+12-\n}{10}\right)}\,,\\
b_3&=\tfrac{\G\left(\tfrac{6}{5}\right)\,\G\left(\tfrac{9}{5}\right)\,\G\left(\tfrac{\n-n^*}{5}\right)}
{\G\left(\tfrac{6+\n}{10}\right)\,\G\left(\tfrac{12+\n}{10}\right)\,\G\left(\tfrac{\n-2n^*+6}{10}\right)}\,.
\end{split}\end{equation}

Having found the explicit solution for $\F(\x)$, we can easily obtain the scaling function by taking
a $\x$-derivative of the former. We prefer not to present the rather unwieldy expression for $\f(\x)$. Instead, it is
more instructive to visualize its functional behavior. In Figure~\ref{fig:2b}, we plot the scaling
function for $p=0.1, 0.5$, and $0.9$. In comparison with lower-$\a$ values, $\f(\x)$ in this case takes a more symmetric
shape, with the persistent weak $p$-dependency.

For larger integer values of the shape function ($\a>3$), exact solutions for $\f(\x)$ could be obtained in principle in a
similar fashion. However, the generalized hypergeometric functions would become increasingly complicated as $\a$ grows.
So, we would not go further in this direction.

\section{Numerical Simulation \label{sec-num}}

In this section, we analyze the stochastic DCS problem by Monte Carlo simulation. Given the generalized product kernel~(\ref{eq:generalized_kernel})--(\ref{eq:generalized_kernel2}),
we develop an algorithm to simulate the binary fragmentation process. Then, we generate
simulated data in order to verify the fractal nature of the system. Our numerical results
match impressively with those predicted by analytical studies. The simulation studies
have been carried out independently in two different programming languages: Python and C++.
The Python codes and the data-sets generated from them are available at \url{https://github.com/rahman-rakib/Stochastic-Fractal}, while those with C++ can be found at
\url{https://github.com/Fahima-Aktar/Stochastic-Fractal}.

What would be an algorithm for simulating the stochastic binary fragmentation process?
For simplicity, we can choose the monodisperse initial condition:
\beq c(x,0)=\d(x-l),\qquad\text{with}\quad l=1.\eeq{num1}
After all, we are interested in the long-time scaling behavior, which ought
to be independent of the initial distribution~\cite{ref.Ziff_McGrady_product}. Now, the fragmentation rate formula~(\ref{frag_rate}) suggests that the splitting probability of
a size-$x$ segment at any given time should go like $x^{2\a-1}$, where $\a$ is the shape
parameter of the symmetric beta distribution that governs the position of breakup points
in accordance with Eq.~(\ref{eq:generalized_kernel2}). Then, the following algorithm makes
sense.
\begin{enumerate}
\item At any iteration step, consider the tuple of surviving fragment lengths:
$(x_1, x_2,\ldots,x_{M_0})$, where $M_0$ is the number of surviving daughter segments.
Construct, for $1\leq j\leq M_0$, the partial sums:
\beq \s_j=\sum_{i=1}^j(x_i)^{2\a-1}.\eeq{num2}
The total sum $\s_{M_0}$$-$identified as the $(2\a\!-\!1)$-th moment $M_{2\a-1}$ of size distribution$-$never exceeds unity for $\a>\tfrac{1}{2}$, thanks to the initial condition~(\ref{num1}).
\item Generate a random number $r_1$ with a uniform distribution in the unit interval $[0,1]$, and compare it with the $(2\a\!-\!1)$-th moment: $M_{2\a-1}\leq1$.
\item
\begin{description}
\item[(\emph{a})] $r_1\leq M_{2\a-1}:$ pick the segment on which $r$ falls, \emph{i.e.}, pick the $j$-th segment such that $\s_j\geq r_1$, but $\s_{j-1}<r_1$, and move on to step 4.
\item[(\emph{b})] $r_1>M_{2\a-1}:$ increase the iteration step by one unit, and go back to step $2$.
\end{description}
\item Generate another random number $r_2\in[0,1]$ with a symmetric beta distribution
of shape parameter $\a$. Break up the length $x_j$ picked in step 3(\emph{a}) into two parts:
$r_2x_j$ and $(1-r_2)x_j$.
\item Generate yet another random number $r_3$ with a uniform distribution in $[0,1]$. If $r_3<p$, keep both the daughter segments produced in step 4. Otherwise, remove one of the daughters, say the one of length $(1-r_2)x_j$, from the system.
\item Increase the iteration step by one unit.
\item Repeat the steps 1 through 6 {\it ad infinitum}.
\end{enumerate}
Steps 1--3(\emph{a}) describe$-$in accordance with the fragmentation
rate~(\ref{frag_rate})$-$how a segment is randomly picked for splitting.
Step 4 spells out where to split the chosen segment; this random event
is modelled by a symmetric beta distribution of shape parameter $\a$,
as the fragmentation kernel~(\ref{eq:generalized_kernel2}) specifies.
Step 5 implements with probability $1-p$ the possible removal of a daughter
segment at any iteration step. Steps 3(\emph{b}) and 6 tells us how to count
the number of iteration steps: it is given by the number of attempts taken
to pick up and subsequently split a segment, regardless of success or failure.

In a stochastic simulation, the measure of time is furnished by the number of
iteration steps:
\beq t = \text{number of iteration steps}.\eeq{num3}
Because of the statistical nature of the problem~\cite{ref.Santo}, it only makes sense
to talk in terms of ensemble averages, which we denote with a pair of angular brackets
as $\langle\,\cdot\,\rangle$. For a given realization in an ensemble, $M_0(t)$ denotes
the total number of segments, whereas $M_1(t)=\sum_{i=1}^{M_0}x_i$ gives the total fragment
length at time $t$. Their respective ensemble averages are denoted as:
\beq N(t)=\langle M_0(t)\rangle,\qquad M(t)=\langle M_1(t)\rangle,\eeq{num4}
while the mean segment size is given by:
\beq \d(t)=M(t)/N(t).\eeq{num5}

In what follows, we work with a $50k$ ensemble size. We start by showing that our
counting of the iteration steps captures time correctly, in agreement with the identification~(\ref{num3}).
To this end, we investigate the long-time behavior of the
mean segment size $\d(t)$.
\begin{figure}[bt]
\centering
\subfloat[]{\includegraphics[width=4.3 cm,
clip=true]{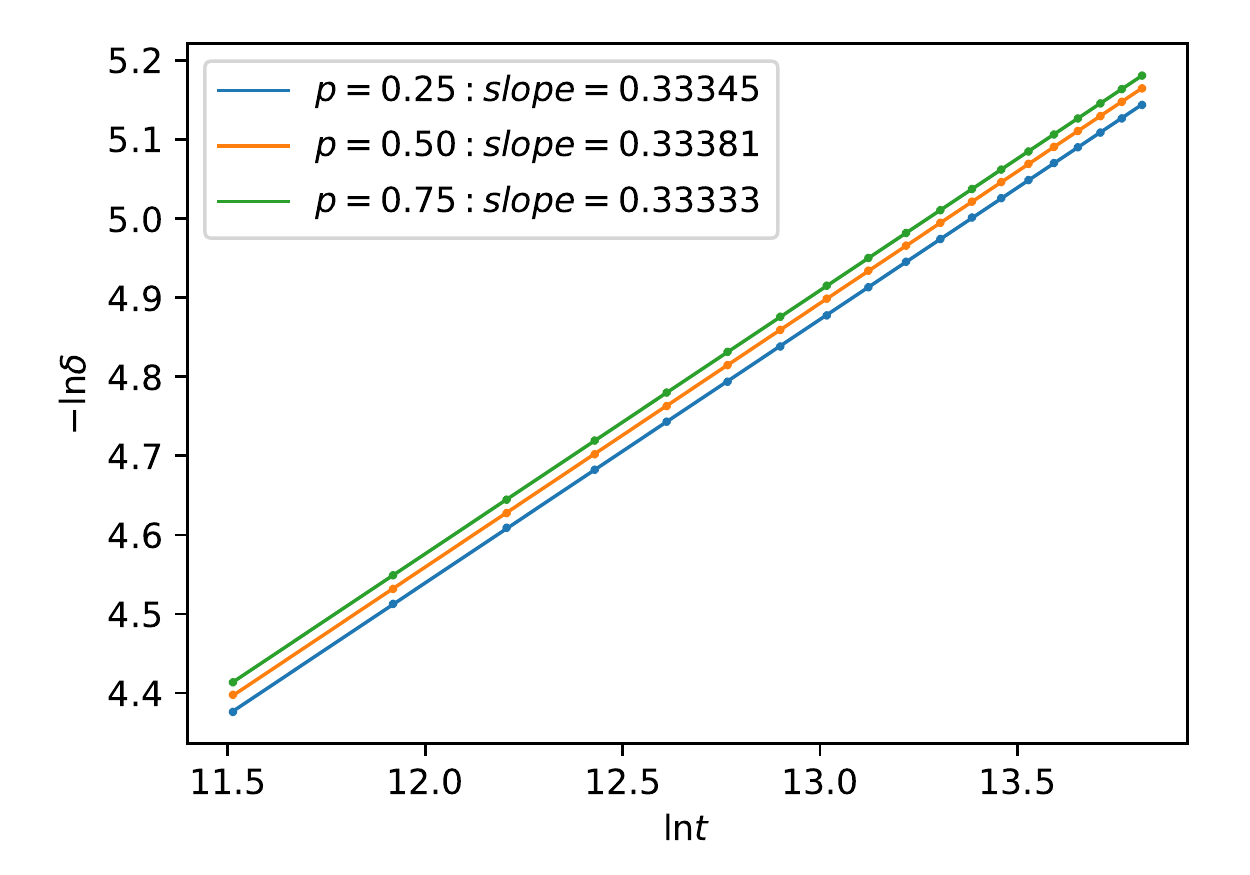} \label{fig:4a}}
\hspace{-10pt}
\subfloat[]{\includegraphics[width=4.3 cm,
clip=true]{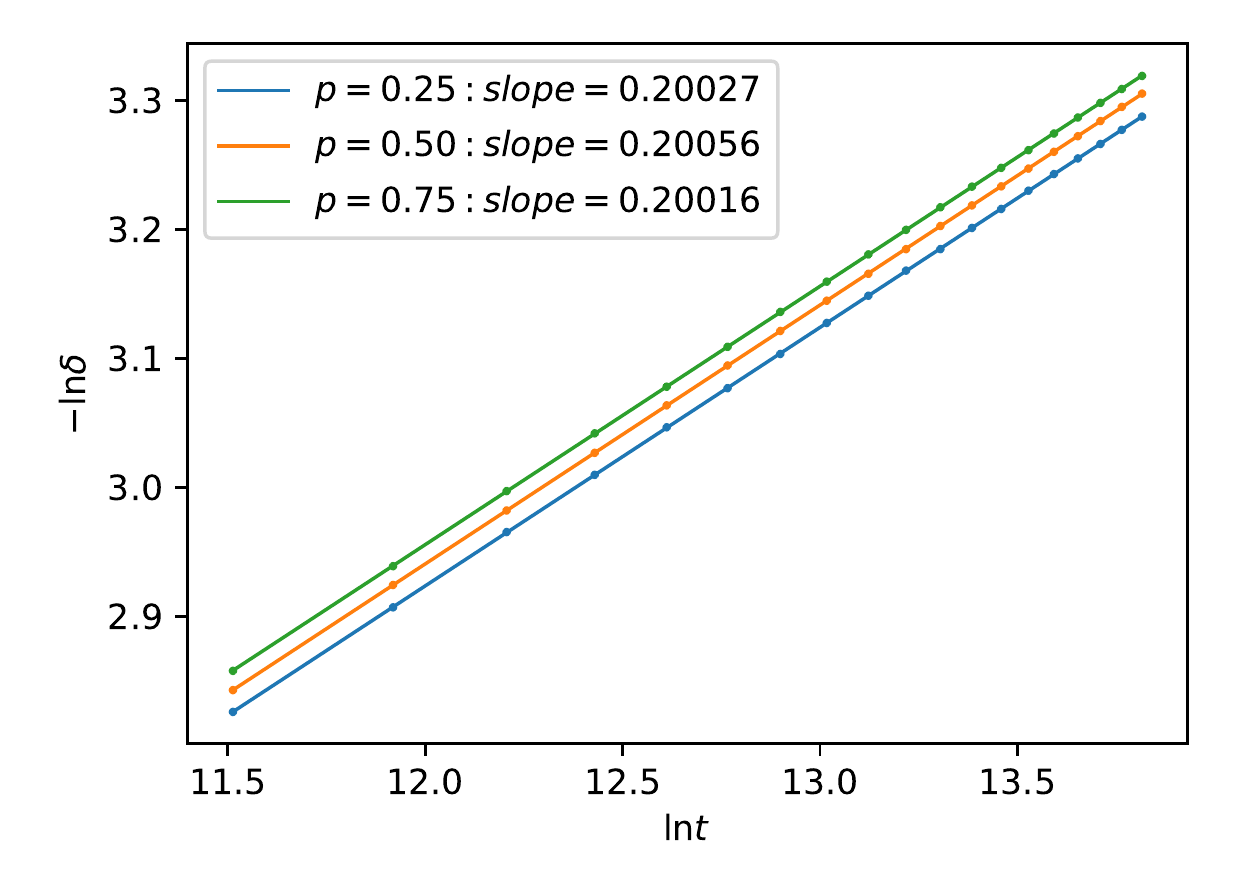} \label{fig:4b}}
\caption{Temporal behavior of the mean segment size $\d$ at large time
for (a) $\a=2$ and (b) $\a=3$.}
\label{fig:4ab}
\end{figure}
In Fig.~\ref{fig:4ab}, we plot $-\ln\d$ versus $\ln t$, at some large time values, $t\in[10^5,10^6]$, for $\a=2,3$.
In each case, our data nicely fits a straight line,
whose slope does not seem to change for different values of probability $p$. In fact,
the slope matches very well with the value $1/(2\a-1)$. This is in total agreement
with the analytical formulation presented in Section~\ref{sec-formulation}. Because
our simulations exhibit the correct scaling behavior~(\ref{scaling-limit}), we must
have identified time $t$ correctly.

\begin{figure}[bt]
\centering
\subfloat[]{\includegraphics[width=4.3 cm,
clip=true]{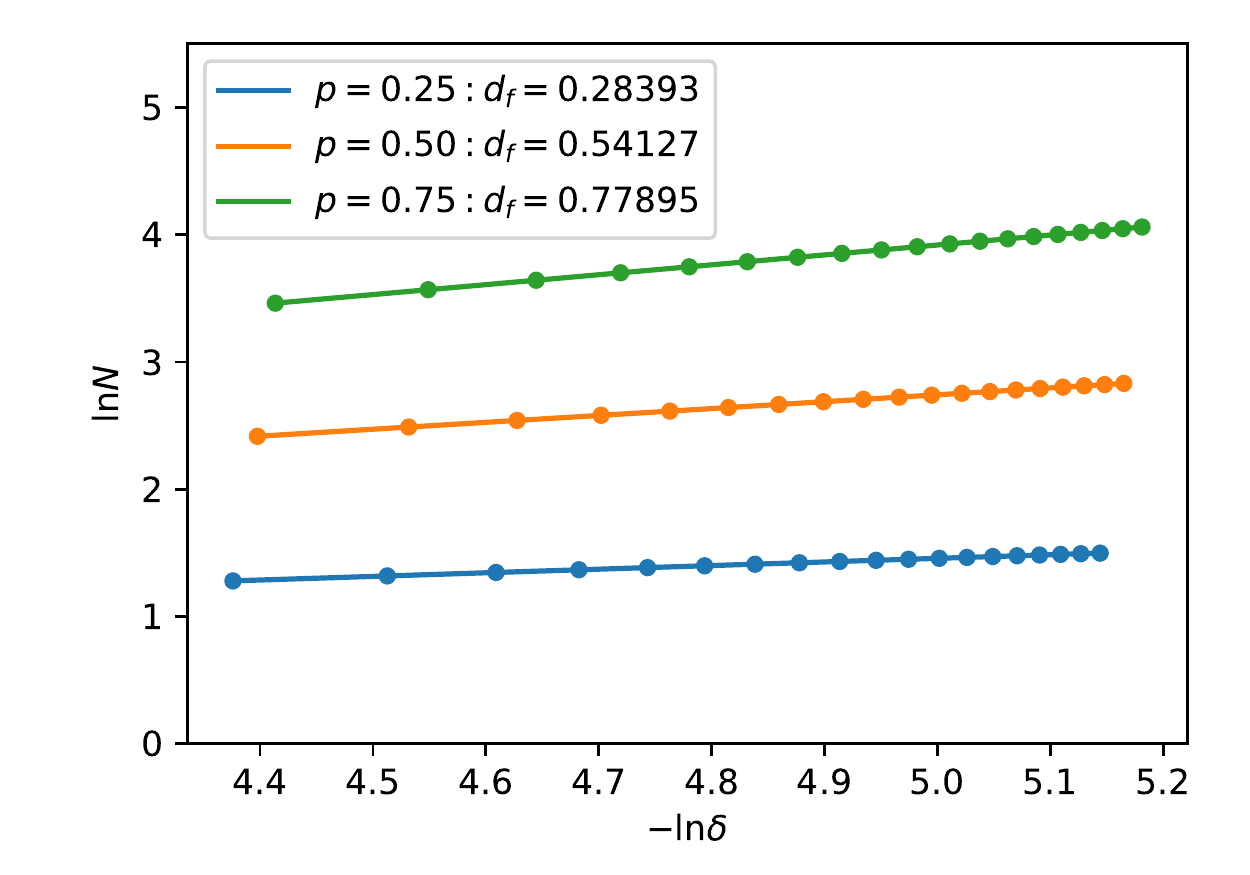} \label{fig:5a}}
\hspace{-10pt}
\subfloat[]{\includegraphics[width=4.3 cm,
clip=true]{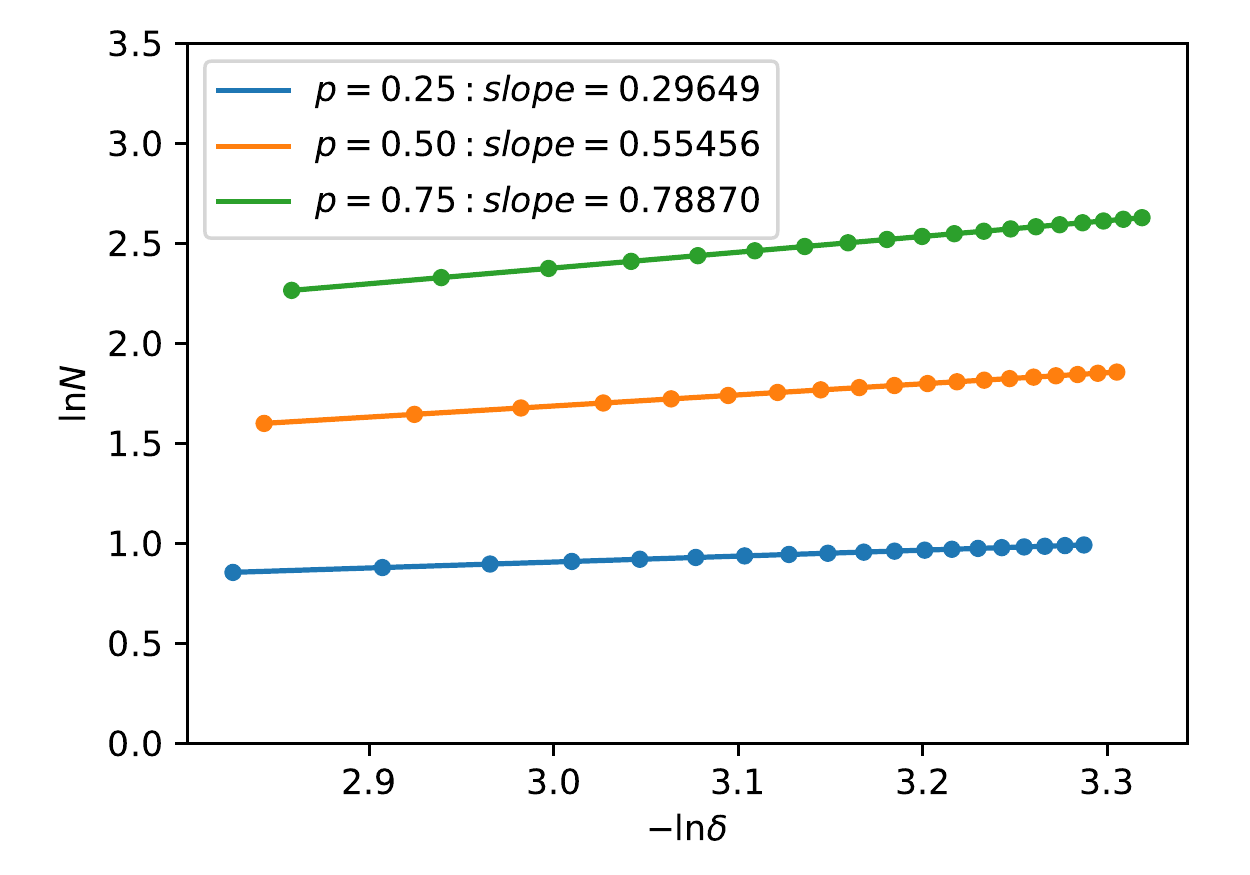} \label{fig:5b}}
\caption{Power-law behavior of the total segment number $N$
w.r.t.~the mean segment size $\d$ for (a) $\a=2$ and (b) $\a=3$.}
\label{fig:5ab}
\end{figure}

Next, we plot $\ln{N}$ versus $-\ln\d$ for given values of $\a$ and $p$, as shown in Fig.~\ref{fig:5ab}.
Again, each plot exhibits linear behavior, and the corresponding
slope is interpreted as the Hausdorff dimension $d_f$ of the system. For different
values of $\a$ and $p$, the slope values match impressively with the respective
$n^*$-values given by Eq.~(\ref{eq:p-nstar}).
This validates the equality of $d_f$ and $n^*$ asserted in Eq.~(\ref{eq:df_obtained}).
The non-integer value $d_f$ quantifies the fractal dimension.

We also compute the $d_f$-th moment of size distribution over the time range
$t\in[10^5,10^6]$, for $\a=2$ and $3$. For a given realization at a given time,
this quantity is obtained by the sum of the $d_f$-th power of all the surviving fragment
\begin{figure}[bt]
\centering
\subfloat[]{\includegraphics[width=4.4 cm,
clip=true]{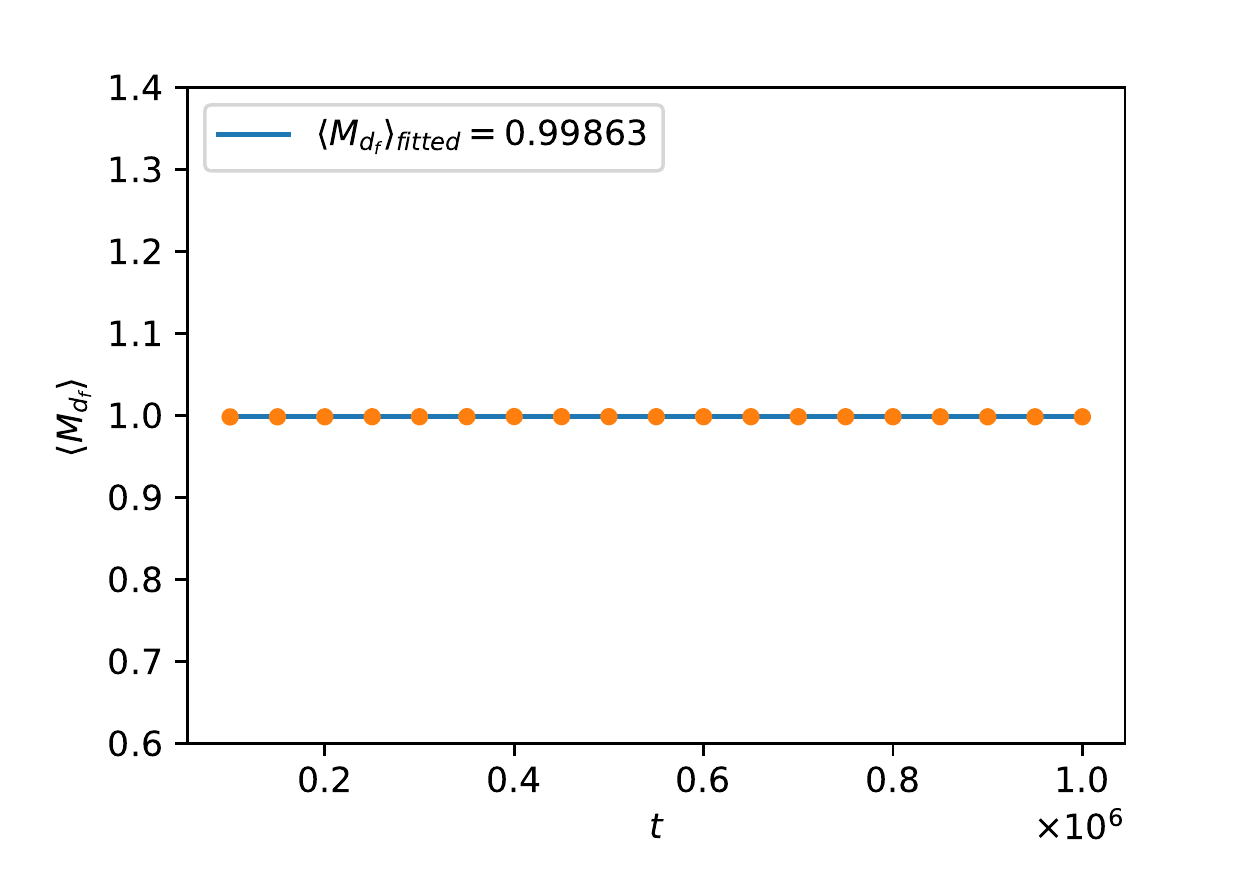} \label{fig:6a}}
\hspace{-15pt}
\subfloat[]{\includegraphics[width=4.4 cm,
clip=true]{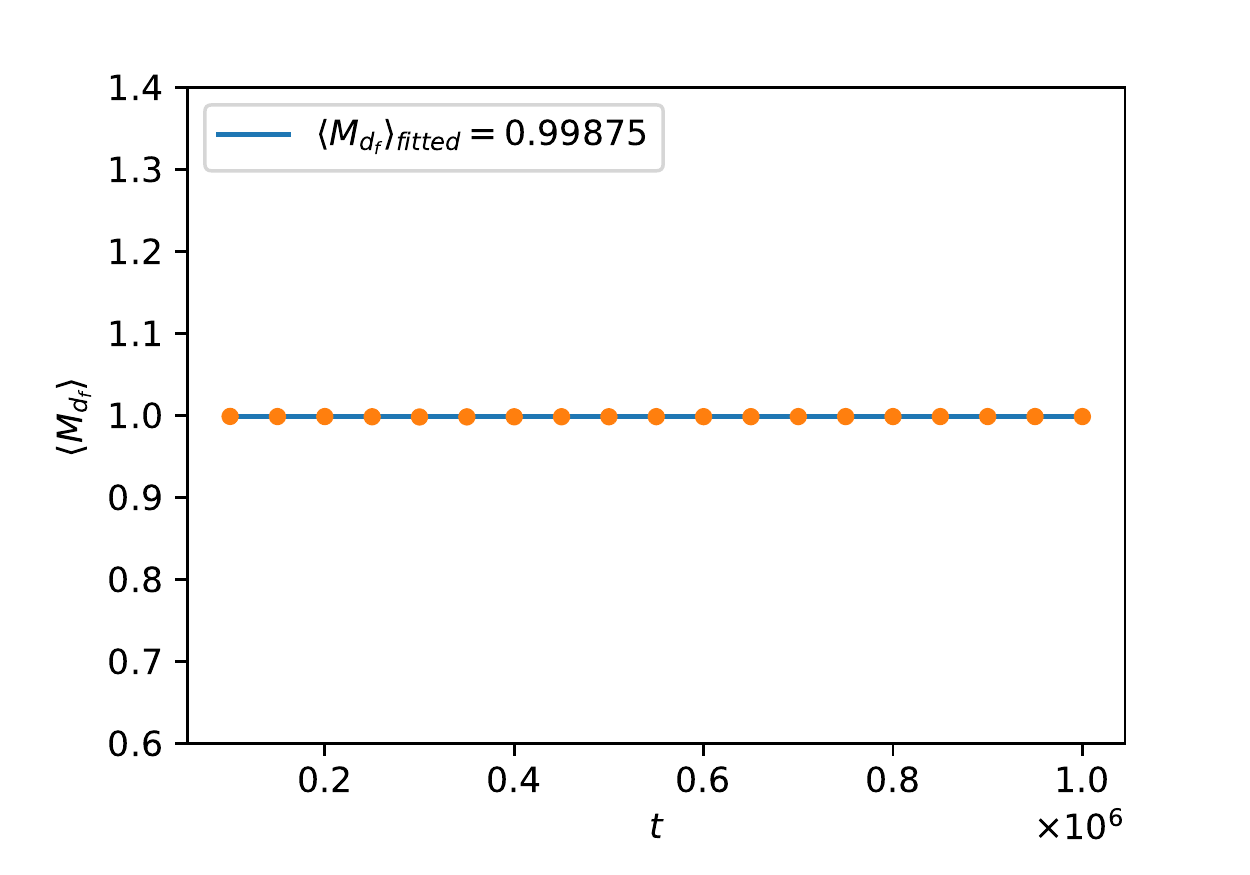} \label{fig:6b}}
\caption{Illustration of the constancy of the ensemble average of $M_{d_f}$
with $p=0.75$ for (a) $\a=2$ and (b) $\a=3$.}
\label{fig:6ab}
\end{figure}
lengths: $M_{d_f}=\sum_{i=1}^{M_0}x_i^{d_f}$. We plot its ensemble average
$\langle M_{d_f}\rangle$ as a function of time $t$, and the results are shown in
Fig.~\ref{fig:6ab}. In each case, the data points have an excellent fit to a horizontal
straight line. This demonstrates that $\langle M_{d_f}\rangle$ is a conserved quantity.
Moreover, for any $\a$, the fitted values of $\langle M_{d_f}\rangle$ equal the theoretical value
of unity~\footnote{It follows from the initial condition~(\ref{num1}) that
$M_{d_f}=1$ at $t=0$. This value remains unchanged for any $t>0$.} with a good accuracy.

Last but not the least, we demonstrate the existence of self-similarity by using the idea of data collapse.
With suitably-chosen bin-widths for $\a=2$ and $3$, we generate ensemble-averaged histogram data that represents
the size distribution $c(x,t)$ as a function of segment-size $x$ at time values: $t=100k, 200k, 300k$, with $p=0.75$.
\begin{figure}[!ht]
\centering
\subfloat[]{\includegraphics[width=4.36 cm,
clip=true]{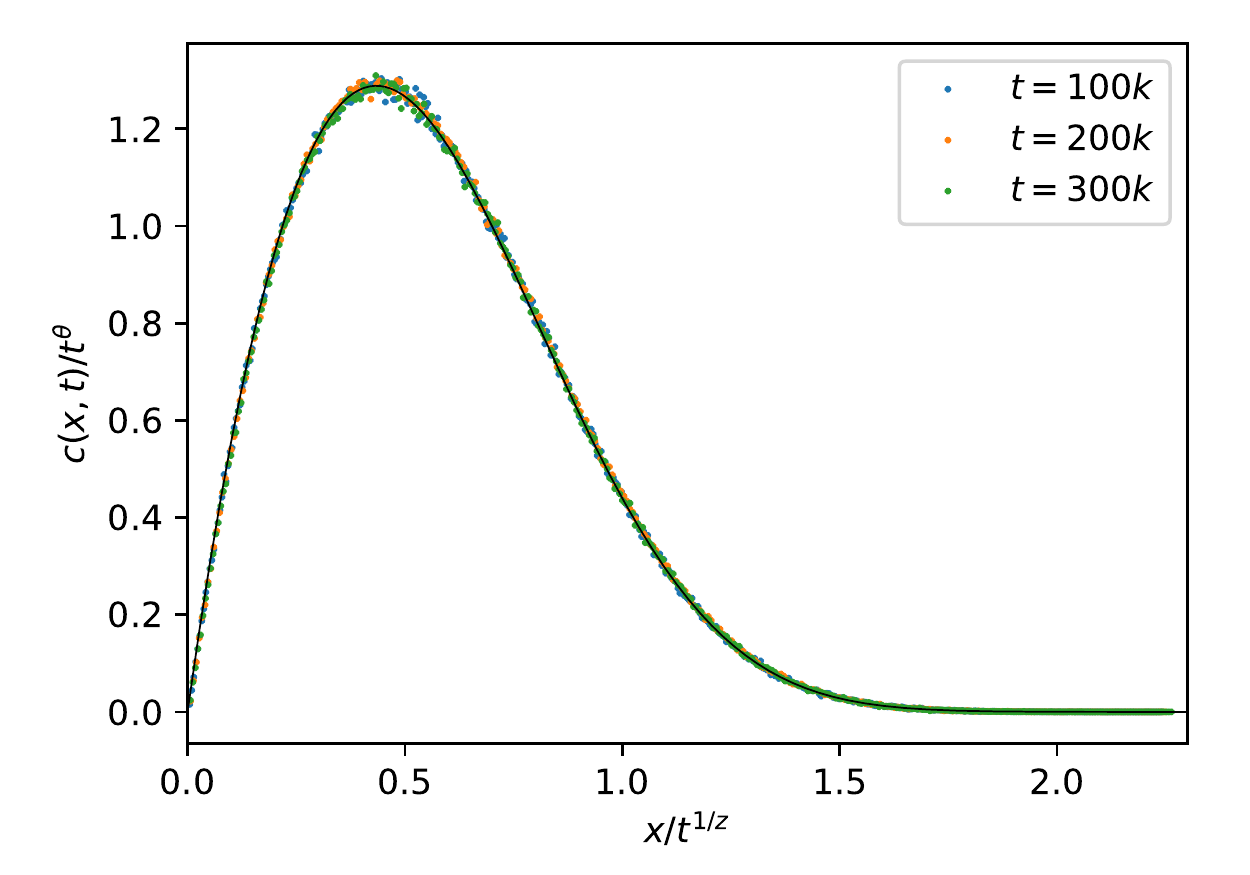} \label{fig:7a}}
\hspace{-11pt}
\subfloat[]{\includegraphics[width=4.36 cm,
clip=true]{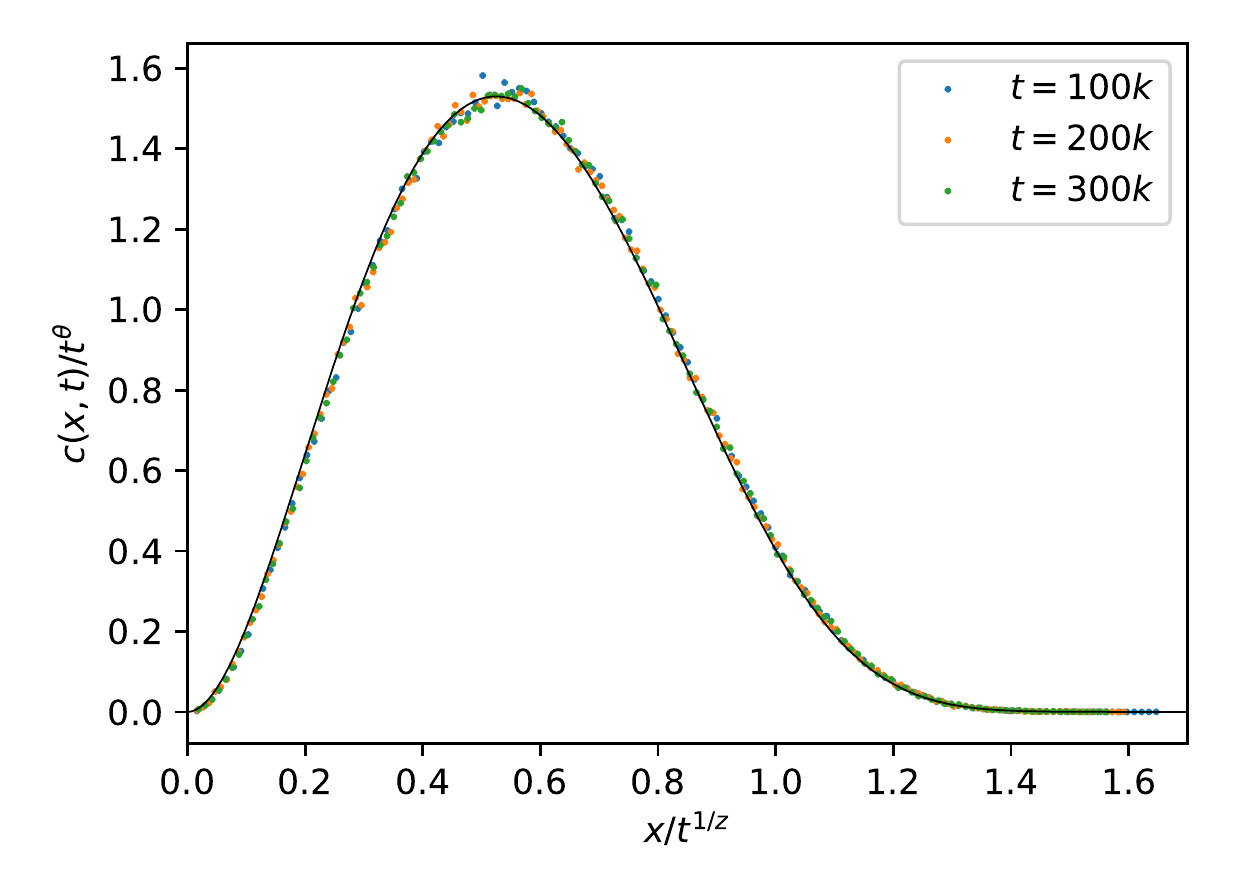} \label{fig:7b}}
\caption{Data collapse plot with $p=0.75$
for (a) $\a=2$ and (b) $\a=3$. The solid line represents analytical solution.}
\label{fig:7ab}
\end{figure}
As already discussed in Section~\ref{expo-sec}, for given $\a$ and $p$, the plots of $c(x,t)/t^\th$ versus $xt^{-1/z}$ at different
$t$-values ought to collapse into a universal curve. Up to a normalization factor, this curve should coincide with the scaling function
$\f(xt^{-1/z})$. As depicted in Fig.~\ref{fig:7ab}, we indeed observe the phenomenon of data collapse by plotting $c(x,t)/t^\th$ against
$xt^{-1/z}$, for the chosen values of $\a$ and $p$. Therein, we have normalized all the data points by an overall factor in order to fit
them with the analytical solutions obtained in Section~\ref{dist-sec}. The fitting to the analytical curves (represented by solid lines)
is excellent, as one can see. This is a strong confirmation of the existence of scaling solutions, which in turn corroborates that the
system exhibits dynamical scaling symmetry$-$the manifestation of self-similarity in a stochastic process.

\section{Symmetry \& Conserved Quantity\label{sec-Noether}}

In the binary fragmentation problem studied in the previous sections, we came across two notions that could be
interconnected by Noether's theorem: a continuous global symmetry and a conserved quantity. On the one hand, the system enjoys
dynamical scaling symmetry that manifests itself through data collapse.
On the other hand, the $d_f$-th moment of the system turns out to be conserved in time. It is then natural
to ask whether the two phenomena are related via Noether's theorem.
In what follows we will resort to the intimate relation between stochastic
processes and Euclidean quantum mechanics (see, for example,~\cite{ref.zambrini, ref.graham_robert} and
references therein). As we will see, one can reinterpret the fragmentation equation~(\ref{eq:binery}) as
the continuity equation in Euclidean time of a nonlinear quantum-mechanical system describing an infinitely-heavy
particle with non-local self interactions.

\subsection{Quantum-Mechanical Description}

In order to find the Schr\"odinger equation of the corresponding quantum-mechanical system, let us note that the
particle size ``$x$'' appearing in Eq.~(\ref{eq:binery}) is a positive semi-definite quantity; we would like to
rename this as ``$r$'', which is to be interpreted as a radial coordinate describing the quantum mechanical system.
Then, the fragmentation equation~(\ref{eq:binery}) reads:
\beq \left[\de_t+a(r)\right]c(r,t)-\int_0^\infty\!\dd r' G(r,r')\,c(r',t)=0,\eeq{st1}
where $a(r)$ is the fragmentation rate defined in Eq.~(\ref{frag_rate_defined}), and the integral corresponds to the gain term, so that
\beq G(r,r')=(1+p)\,\th(r'-r)\,F(r'-r,r),\eeq{st2}
with $\th(r-r')$ being the Heaviside step function:
\beq \th(r-r')=\left\{
                 \begin{array}{ll}
                   0, & \hbox{$r<r'$;} \\
                   1, & \hbox{$r\geq r'$.}
                 \end{array}
               \right.
\eeq{st3}

Because the particle number density is real and positive semi-definite:
$c(r,t)\geq0$, it can be expressed as:
\beq c(r,t)=|\Ps(r,t)|^2,\eeq{st4}
where $\Ps(r,t)$ is some complex function to be determined. To proceed, we make the
following ansatz for the equation that governs the latter function:
\beq \left[\de_t+V(r,t)\right]\Ps(r,t)=0,\eeq{st5}
where the function $V(r,t)$ is assumed to be real-valued. To see that the
ansatz~(\ref{st5}) qualifies, let us multiply it by $\Ps^*(r,t)$
and also take complex conjugate of the resulting equation. The sum of
the two gives:
\beq \left[\de_t+2V(r,t)\right]|\Ps(r,t)|^2=0.\eeq{st6}
In view of the identification~(\ref{st4}), one can compare Eqs.~(\ref{st1}) and~(\ref{st6})
to obtain the following result:
\beq V(r,t)=\tfrac{1}{2}a(r)-\tfrac{1}{2}\!\int_0^\infty\!\dd r' G(r,r')
\biggl|\frac{\Ps(r',t)}{\Ps(r,t)}\biggr|^2.\eeq{st7}

Let us make an analytic continuation in time:
\beq t=i\t.\eeq{st8}
Then, using the notation: $\ps(r,\t)=\Ps(r,i\t)$, one finds that Eq.~(\ref{st5}) gives rise to the following
form:
\beq i\de_\t\ps(r,\t)=\hat{H}(r,\t)\ps(r,\t),\eeq{st9}
for some $\hat{H}(r,\t)$ to be specified shortly. With $\hbar=1$, we would like to interpret Eq.~(\ref{st9}) as
a Schr\"odinger equation in the radial coordinate. The ``Hamiltonian'' $\hat{H}(r,\t)$ pertains to an infinitely-heavy
particle, say in dimensions $D$, with non-local self interactions (see~\cite{ref.weinberg} for a systematic study of
nonlinear quantum mechanics). Explicitly,
\beq \hat{H}(r,\t)=\lim_{m\rightarrow\infty}\left[-\tfrac{1}{2m}\left(\de^2_r+\tfrac{D-1}{r}\de_r\right)\right]
+\hat{V}(r,\t),\eeq{st10}
where the potential $\hat{V}(r,\t)=V(r,i\t)$ is real-valued; it consists of two parts: a local potential
$\hat{V}_0(r)$ and a non-local self interaction term $\hat{V}_1(r,\t)$, given as:
\begin{equation}\label{st11} \begin{split}
&\hat{V}(r,\t)=\hat{V}_0(r)+\hat{V}_1(r,\t),\\
\hat{V}_0=\tfrac{1}{2}a(r),\quad &\hat{V}_1=-\tfrac{1}{2}\!\int_0^\infty\!\dd r' G(r,r')
\biggl|\frac{\ps(r',\t)}{\ps(r,\t)}\biggr|^2.\end{split}\end{equation}
Upon inspection, it is clear that the local potential $\hat{V}_0$ corresponds to the loss term in the
fragmentation equation~(\ref{st1}), whereas the non-local self interaction $\hat{V}_1$ to the gain term.
The nonlinear Schr\"odinger equation described by Eqs.~(\ref{st9})--(\ref{st11}) respects the ``homogeneity''
property~\cite{ref.weinberg}, according to which if $\ps$ is a solution, then so is $Z\ps$
for an arbitrary complex constant $Z$. This fact will play an important role in your subsequent analysis.

The wave function $\ps(r,\t)$ appearing in Eq.~(\ref{st9}) is a complex Schr\"odinger field. Its complex conjugate
$\ps^*(r,\t)$ obeys the equation of motion:
\beq -i\de_\t\ps^*(r,\t)=\hat{H}(r,\t)\ps^*(r,\t).\eeq{st12}
Can Eqs.~(\ref{st9}) and~(\ref{st12}) be derived from an action through variational principle? The answer is no$-$the
equations as such are non-Lagrangian~\footnote{This can be seen by noting that a Lagrangian may exist only if the two functional derivatives:
$\tfrac{\de\mathcal{R}}{\de\ps^*}$ and $\tfrac{\de\mathcal{R}^*}{\de\ps}$ are equal, where $\mathcal{R}$ ($\mathcal{R}^*$) is
the equation of motion of $\ps$ ($\ps^*$). The nonlocal interaction term violates this condition.}. However, an action exists if the self interactions are excluded.
Assuming spherical symmetry, upon integrating out the angular coordinate(s), the action can be written, up to a factor, as:
\beq S=\int_0^\infty\!\dd \t\int_0^\infty\!\dd r\,\mathcal{L}[\ps_0,\ps_0^*],\eeq{st13}
where the subscript ``$0$'' indicates the exclusion of self interactions, and $\mathcal{L}$
is the Lagrangian density:
\beq \mathcal{L}=r^{D_0-1}\left[\tfrac{i}{2}\left(\ps_0^*\dot\ps_0-\dot\ps_0^*\ps_0\right)
-\hat{V}_0(r)\ps_0^*\ps_0\right],\eeq{st14}
with a ``dot'' denoting a derivative w.r.t.~time $\t$. Note that the dimension seen by the quantum-mechanical system in the absence
of self interactions is denoted by $D_0$, instead of $D$; this distinction can be justified \emph{a posteriori}.
While the action~(\ref{st13})--(\ref{st14}) does not correspond to our original problem, it will still provide valuable
guidance on constructing conserved charges of the self-interacting quantum-mechanical system we are interested in.

\subsection{Global Symmetries and Noether Charges}\label{sec:Charge}

In this section, we will explore continuous global symmetries of the equations of
motion~(\ref{st9}) and~(\ref{st12}). Conventionally, Noether's theorem is presented in
the context of symmetries of an action. However, as noted in~\cite{ref.kaparulin}, the
Lagrangian formulation of a system is not essential in order for connecting continuous
global symmetries with conserved quantities. The latter point is particularly important
given the non-Lagrangian nature of our quantum-mechanical system. Moreover, Noether's
theorem can aptly be generalized for non-local theories~\cite{ref.oriti}.

To find the desired Noether charges, we proceed as follows. For a given symmetry, we
first consider the system without self interactions $\hat{V}_1$, which enjoys the Lagrangian
formulation~(\ref{st13})--(\ref{st14}). This enables us to compute the Noether charge from
the textbook formula:
\beq Q_0=\int_0^\infty\!\dd r\left(\frac{\de\mathcal L}{\de\dot\ps_0}\d_{\e}\ps_0
+\frac{\de\mathcal L}{\de\dot\ps_0^*}\d_{\e}\ps_0^*+\mathcal{L}\de_{\e}\t\right),\eeq{ph0}
where $\d_{\e}$ denotes variation w.r.t.~an infinitesimal symmetry transformation parameter
$\e$. With the knowledge of $Q_0$, we make an educated guess at the conserved charge $Q$ when
the self interactions are included back. For any trial charge $Q$, we can use the equations of
motion~(\ref{st9}) and~(\ref{st12}) to check if it is indeed conserved.

Because constants of motion do \emph{not} depend on time, they continue to be conserved even if
time is analytically continued to imaginary values. This seemingly na\"ive fact actually
plays a crucial role. As we will see, the symmetries and conserved charges of the binary fragmentation
problem~(\ref{eq:binery}) may be connected only in Euclidean time.

\subsubsection{Phase Rotation}

The equations of motion~(\ref{st9}) and~(\ref{st12}) are invariant under the global phase rotation:
\beq \ps\rightarrow e^{-i\tilde\a}\ps,\qquad \ps^*\rightarrow e^{+i\tilde\a}\ps^*,\eeq{ph1}
where $\tilde\a$ is a constant phase angle. This symmetry simply reflects the homogeneity property
of these equations noted earlier. With self interactions excluded, the infinitesimal
variations read: $\d_{\e}\ps_0=-i\ps_0$, $\d_{\e}\ps_0^*=+i\ps^*$, and $\d_{\e}\t=0$. The corresponding
Noether charge is given by formula~(\ref{ph0}) as:
\beq Q_{0,p}=\int_0^\infty\!\dd r\,r^{D_0-1}|\ps_0(r,\t)|^2.\eeq{ph2}
In the context of Quantum Mechanics, the charge $Q_{0,p}$ reflects nothing but the conservation of
probability. Upon including self interactions, let us try with the charge:
\beq Q_{p}=\int_0^\infty\!\dd r\,r^{D-1}|\ps(r,\t)|^2.\eeq{ph3}
Its time-derivative is given by the integral:
\beq \dot{Q}_{p}=\int_0^\infty\!\dd r\,r^{D-1}\left(\ps^*\dot{\ps}+\ps\dot{\ps}^*\right),\eeq{ph4}
which clearly vanishes on account of the equations of motion~(\ref{st9}) and~(\ref{st12}). Therefore,
the $Q_{p}$ given in Eq.~(\ref{ph3}) is the correct conserved charge.

Now, because $Q_{p}$ is time-independent, we can Wick rotate $\t$ back to $t$ to get a charge conservation:
\beq \mathcal{Q}_{p}=\int_0^\infty\!\dd r\,r^{D-1}\,c(r,t),\quad \text{with}\quad
\frac{d}{dt}\mathcal{Q}_{p}=0.\eeq{ph5}
Therefore, the $(D\!-\!1)$-th moment of the system undergoing binary fragmentation must be
conserved in time. The results of Section~\ref{sec-gen}, then, lead us to the identification:
\beq D = d_f+1.\eeq{ph6}
The $d_f$-th moment, as a conserved charge, has therefore nothing to do with dynamical scaling.
Moreover, quite remarkably, the symmetry associated with this conservation does not even ``exist''
in real time.

One can turn the logic around to make nontrivial conclusions for the quantum-mechanical system.
The presence of non-local self interactions in Eqs.~(\ref{st9})--(\ref{st11}),
for $0<p<1$, invokes fractional dimension. This is reminiscent of the interplay between non-locality
and fractional dimension observed for fractional Schr\"odinger equation~\cite{ref.jeng,ref.wei,ref.modanese}.
In the definition~(\ref{ph3}) of $Q_p$, if one identifies the dimension $D$ with the integer dimension
$D_0$ that appears in the absence of self interactions, one ends up with probability non-conservation.
However, this conclusion is na\"ive. As Eq.~(\ref{ph6}) suggests, the fractional nature of space dimensions
must be taken into due account. Then, probability conservation indeed holds.

\subsubsection{Dynamical Scaling}

The $d_f$-th moment conservation, as we have already understood, is not connected with dynamical scaling.
Then, what is the conserved quantity corresponding to this continuous global symmetry? To answer
this question, let us first recall from Eqs.~(\ref{dscaling1})--(\ref{dscaling2}) that the symmetry transformations are implemented as:
\beq r\rightarrow \l\,r,\qquad \t\rightarrow \l^z\t;\qquad z=-(\D+1),\eeq{ds1}
for some $\l\in\mathbb{R}_+$. Indeed, Eqs.~(\ref{st9}) and~(\ref{st12}) are invariant
under the transformation~(\ref{ds1}) given that the wave function $\ps$ and its complex
conjugate $\ps^*$ scale as:
\beq \ps\rightarrow \l^{z\th/2}\ps,\qquad \ps^*\rightarrow \l^{z\th/2}\ps^*.\eeq{ds2}

The value of the exponent $\th$ can be reconfirmed by noting that a Noether charge should have a zero scaling
dimension. For example, the conserved charge $Q_{p}$, given in Eq.~(\ref{ph3}), by definition does not depend on time.
Neither can it be a function of the space coordinate, which is integrated out. Then, invariance of $Q_p$ gives:
\beq \th=\tfrac{D}{\D+1}.\,\eeq{ds4}
Given the identification~(\ref{ph6}), this is of course in complete agreement with the results obtained
Section~\ref{sec-gen} by assuming the existence of scaling solutions for $z<0$.

When self interactions are excluded, one may consider the invariance of the action~(\ref{st13})--(\ref{st14})
under dynamical scaling. This gives a similar expression for $\th$, with the substitution:
$D\rightarrow D_0$ in Eq.~(\ref{ds4}). In this case, an infinitesimal transformation with $\l = 1+\e$ gives:
\beq \d_\e \t=z\t,\quad \d_\e r=r,\quad \d_\e\ps=\tfrac{1}{2}z\th_0\ps,\quad \d_\e\ps_0^*=\tfrac{1}{2}z\th_0\ps_0^*.\eeq{ds5}
Upon using formula~(\ref{ph0}), and the fact that the Lagrangian vanishes on the equations
of motion, the corresponding charge is found to be trivial:
\beq Q_{0,s}=0.\eeq{ds6}

Now, we will show that the Noether charge remains trivial even in the presence of self interactions.
Note that a nontrivial charge  would be an integral of the form:
\beq Q^{(trial)}_{s}=\int_0^\infty\!\dd{r}f(r)\int_0^\infty\!\dd r' G(r,r')|\ps(r',\t)|^2,\eeq{ds7}
with some yet-unspecified function $f(r)\neq0$. It is clear from Eqs.~(\ref{st11}) that the trial charge~(\ref{ds7}) vanishes
in the absence of self-interaction terms, as it should. Moreover, its time derivative is zero: $\dot{Q}^{(trial)}_{s}=0$,
irrespective of $f(r)$. This is because $\tfrac{d}{d\t}|\ps(r',\t)|^2$ itself vanishes on the equations of motion~(\ref{st9})
and~(\ref{st12}). In congruence with the invariance under dynamical scaling of the charge~(\ref{ds7}) itself, we can choose:
$f(r)=r^{D-\D-2}\neq0$.

Again, because $Q^{(trial)}_{s}$ is time-independent, we can Wick rotate $\t$ back to $t$, and still get a charge:
\beq \mathcal{Q}^{(trial)}_s=\int_0^\infty\!\dd r\,r^{D-\D-2}\!\int_0^\infty\!dr'G(r,r')c(r',t),\eeq{ds10}
which is conserved in time $t$. The charge appearing in Eq.~(\ref{ds10}) is just a Mellin transform
of the gain term appearing in the fragmentation equation~(\ref{st1}). To get a better insight into this quantity, let us
take an appropriate Mellin transform of Eq.~(\ref{st1}) to obtain:
\beq \frac{d}{dt}M_{d_f-\D-1}(t)=-\mathcal{Q}_p+\mathcal{Q}^{(trial)}_s,\eeq{ds11}
where we have made use of the identification~(\ref{ph6}) and the choice $a(r)=r^{\D+1}$. The left-hand side of Eq.~(\ref{ds11})
is however proportional to $\mathcal{Q}_p$ itself. This can be seen from the rate equation~(\ref{eq:rate_equation})
obtained for a generalized product kernel with shape parameter $\a=1+\D/2$. The upshot is the linear dependency of two charges:
\beq \mathcal{Q}^{(trial)}_s=\left(1+\g_{d_f-2\a+1}\right)\mathcal{Q}_p\,.\eeq{ds12}
This means that, for $f(r)\neq0$, the proposed charge~(\ref{ds7}) corresponds instead to phase rotation.
The conclusion is that $f(r)$ cannot be nontrivial if the charge~(\ref{ds7}) has to do with dynamical scaling.
This means:
\beq \mathcal{Q}_s=0.\eeq{trivial_charge}
In other words, there is no nontrivial Noether charge corresponding to dynamical scaling.

\section{Remarks \label{sec-conc}}

In this article, we have studied the formation of a stochastic fractal by invoking
a stochastic version of the dyadic Cantor set problem. The process entails that,
at any step, only one segment may split$-$at a random point, preferentially
w.r.t.~its size$-$into two fragments. One of the daughter segments may immediately
disappear from the system with probability $1-p$. For a continuous system, such
kinetics can be captured by a variant of the well-known binary fragmentation equation.
We have considered a fragmentation kernel, for which segment breakup points follow
a symmetric beta distribution with shape parameter $\a>\tfrac{1}{2}$, while the
fragmentation rate goes like the $(2\a\!-\!1)$-th power of segment-size. We have
demonstrated$-$both analytically and through extensive Monte Carlo simulations$-$that
the system exhibits fractal properties in the long-time limit. This was established
by two important observations: the emergence of a fractal dimension, and the
manifestation of self-similarity through dynamical scaling symmetry.

The shape parameter $\a$ encodes randomness in breaking segments preferentially
in the center. The present work generalizes the study~\cite{ref.Pandit} of the
completely random case ($\a=1$). As the degree of this randomness decreases, the fractal
dimension is found to increase, reaching the maximum value $\log_2(1+p)$ in the
no-randomness limit: $\a\rightarrow\infty$. Our analytical results include the
prediction of a conserved moment for generic values of $\a$ and $p$. We have also
presented analytical expressions for the scaling forms of size distribution in the
cases of $\a=2$ and $3$. On the other hand, our algorithm for simulating the
stochastic process works nicely for arbitrary $\a$ and $p$. This is a nontrivial
feat to achieve, especially because it is \emph{a priori} not obvious at all how
to correctly incorporate time in simulation studies for a non-uniform distribution
$(\a\neq1)$. It is reassuring to find excellent agreement between
the analytical and the simulation-based results.

The deep connection between stochastic processes and Euclidean quantum mechanics
can hardly be overstated. Indeed, our exploration of the symmetry
origin of the conserved quantity relies on it. Corresponding to the binary fragmentation process,
there appears in Euclidean time a quantum-mechanical system of an infinitely-heavy particle with
non-local self interactions. We have thereby revealed that dynamical scaling symmetry does not
give rise to any nontrivial conservation law. Remarkably, the symmetry
associated with the conserved moment turns out to be purely mathematical in nature: phase rotation
in the quantum-mechanical system, which does not even ``exist'' in real time~\footnote{This is reminiscent
of the Laplace--Runge--Lenz vector in classical mechanics, which appears as a constant of motion in the
Kepler problem. Its conservation, however, corresponds to a purely mathematical symmetry.}. Therefore, the nontrivial
conservation law in the binary fragmentation problem parallels probability conservation in quantum mechanics.
This connection could have important implications for quantum mechanical systems in the presence of non-locality.

It would be interesting to consider the opposite phenomenon of fragmentation, namely \emph{kinetics of aggregation}.
Fractal behavior and conservation laws have been noticed in such processes by including stochastic
self-replication with certain probability and constant kernel~\cite{ref.hassan_nabila}.
One could explore this direction further, for generalized kernels, and study the corresponding Euclidean quantum-mechanical
systems. Another avenue to pursue is the generalization
of the present work to dimensions larger than one. For $p=1$, higher-dimensional fragmentation processes exhibit
multi-scaling with infinitely many conserved quantities~\cite{ref.k_b-n}, which in turn give rise to
multi-fractality~\cite{ref.hassanRogers2,ref.hassan,ref.hassan_dayeen}.
It is \emph{a priori} unclear what could furnish the corresponding infinitude of
continuous global symmetries. We leave these as future work.


\end{document}